\documentclass[numberedappendix]{emulateapj}

\newcommand{\cote}{C\^{o}t\'{e}\ }
\newcommand{\jordan}{Jord\'{a}n\ }
\newcommand{\hasegan}{Ha\c{s}egan\ }
\newcommand{\etal}{et~al.\ }

\newcommand{\sersic}{S\'ersic}

\slugcomment{Accepted for publication in The Astrophysical Journal}

\shorttitle{Intergalactic Globular Clusters in the Coma Cluster}
\shortauthors{Peng et al.}

\begin{document}


\title{The {\it HST/ACS} Coma Cluster Survey IV.  Intergalactic
  Globular Clusters and the Massive Globular
  Cluster System at the Core of the Coma Galaxy Cluster
  \altaffilmark{1}}


\author{Eric W. Peng\altaffilmark{2,3,4}}

\author{Henry C. Ferguson\altaffilmark{4}}

\author{Paul Goudfrooij\altaffilmark{4}}

\author{Derek Hammer\altaffilmark{5}}

\author{John R. Lucey\altaffilmark{6}}

\author{Ronald O. Marzke\altaffilmark{7}}

\author{Thomas H. Puzia\altaffilmark{8,9}}

\author{David Carter\altaffilmark{10}}

\author{Marc Balcells\altaffilmark{11}}

\author{Terry Bridges\altaffilmark{12}}

\author{Kristin Chiboucas\altaffilmark{13}}

\author{Carlos del Burgo\altaffilmark{14}}

\author{Alister W. Graham\altaffilmark{15}}

\author{Rafael Guzm\'an\altaffilmark{16}}

\author{Michael J. Hudson\altaffilmark{17}}

\author{Ana Matkovi\'c\altaffilmark{18}}

\author{David Merritt\altaffilmark{19}}

\author{Bryan W. Miller\altaffilmark{20}}

\author{Mustapha Mouhcine\altaffilmark{10}}

\author{Steven Phillipps\altaffilmark{21}}

\author{Ray Sharples\altaffilmark{6}}

\author{Russell J. Smith\altaffilmark{6}}

\author{Brent Tully\altaffilmark{12}}

\author{Gijs Verdoes Kleijn\altaffilmark{22}}



\altaffiltext{1}{Based on observations with the NASA/ESA {\it Hubble
    Space Telescope} obtained at the Space Telescope Science Institute,
  which is operated by the Association of Universities for Research in
  Astronomy, Inc., under NASA contract NAS 5-26555.}
\altaffiltext{2}{Department of Astronomy, Peking University, Beijing
  100871, China; peng@bac.pku.edu.cn}
\altaffiltext{3}{Kavli Institute for Astronomy and Astrophysics, Peking
  University, Beijing 100871, China}
\altaffiltext{4}{Space Telescope Science Institute,
  3700 San Martin Drive, Baltimore, MD 21228, USA}
\altaffiltext{5}{Department of Physics and Astronomy, Johns Hopkins University
  3400 N. Charles St., Baltimore, MD 21228, USA}
\altaffiltext{6}{Department of Physics, University of Durham, South Road, Durham
  DH1 3LE, UK}
\altaffiltext{7}{Dept. of Physics \& Astronomy, San Francisco State University,
  1600 Holloway Avenue, San Francisco, CA 94132, USA}
\altaffiltext{8}{National Research Council of Canada,
  Herzberg Institute of Astrophysics,    
  5071 West Saanich Road, Victoria, BC  V9E 2E7, Canada} 
\altaffiltext{9}{Departmento de Astronom\'ia y Astrof\'isica,
  Pontificia Universidad Cat\'olica de Chile, Av.\ Vicu\~na Mackenna
  4860, 7820436 Macul, Santiago, Chile }
\altaffiltext{10}{Astrophysics Research Institute, Liverpool John
  Moores University, Twelve Quays House, Egerton Wharf, Birkenhead
  CH41 1LD, UK}
\altaffiltext{11}{Instituto de Astrofisica de Canarias, 38200 La
  Laguna, Tenerife, Spain}
\altaffiltext{12}{Department of Physics, Engineering Physics and Astronomy,
Queen's University, Kingston, ON K7L 3N6, Canada}
\altaffiltext{13}{Institute for Astronomy, University of Hawai'i, 2680 Woodlawn
  Drive, Honolulu, HI 96822, USA}
\altaffiltext{14}{School of Cosmic Physics, Dublin Institute for Advanced
Studies, 31 Fitzwilliam Place, Dublin 2, Ireland}
\altaffiltext{15}{Centre for Astrophysics and Supercomputing, Swinburne
  University of Technology, Hawthorn, Victoria 3122, Australia}
\altaffiltext{16}{Department of Astronomy, University of Florida, P.O. Box
  112055, Gainesville, FL 32611, USA}
\altaffiltext{17}{Physics and Astronomy, University of Waterloo, 200 University
  Avenue West, Waterloo, Ontario, Canada N2L 3G1, Canada}
\altaffiltext{18}{Department of Astronomy and Astrophysics, Pennsylvania State University, University Park, PA 16802, USA}
\altaffiltext{19}{Center for Computational Relativity and Gravitation and
  Department of Physics, Rochester Institute of Technology, Rochester,
  NY 14623, USA}
\altaffiltext{20}{Gemini Observatory, Casilla 603, La Serena, Chile}
\altaffiltext{21}{Astrophysics Group, H. H. Wills Physics Laboratory, University
  of Bristol, Tyndall Avenue, Bristol BS8 1TL, UK}
\altaffiltext{22}{Kapteyn Astronomical Institute, University of Groningen,
  P.O. Box 800, 9700 AV Groningen, The Netherlands}


\begin{abstract}
Intracluster stellar populations are a natural result of
tidal interactions in galaxy clusters.  Measuring these
populations is difficult, but important for understanding the assembly of
the most massive galaxies.
The Coma cluster of galaxies is one of the nearest truly massive galaxy
clusters, and is host to a correspondingly large system of globular
clusters (GCs).  We use imaging from the {\it HST/ACS} Coma Cluster
Survey to present the first definitive detection of a large population of
intracluster GCs (IGCs) that fills the Coma cluster core and is not
associated with individual galaxies.  The GC surface density
profile around the central massive elliptical galaxy, NGC~4874, is
dominated at large 
radii by a population of IGCs that extend to the limit of our data
($R<520$~kpc).  We 
estimate that there are $47000\pm1600$ (random)
$^{+4000}_{-5000}$ (systematic) IGCs out to this radius, and that they
make up $\sim70\%$ of the central GC system, making this the
largest GC system in the nearby Universe.  Even including the GC
systems of other cluster galaxies, the IGCs still make up
$\sim30$--$45\%$ of the GCs in the cluster core.  Observational limits
from previous studies of the intracluster light (ICL) suggest that the IGC
population has a high specific frequency. If the IGC population has a
specific frequency similar to high-$S_N$ dwarf galaxies, then the ICL
has a mean surface brightness of $\mu_V\approx 27 {\rm\ mag\
  arcsec}^{-2}$ and a total stellar mass  
of roughly $10^{12} \mathcal{M}_\odot$ within the cluster core.  The ICL makes
up approximately half of the stellar luminosity and one-third of the
stellar mass of the central (NGC4874+ICL) system.  The
color distribution of the IGC population is bimodal,
with blue, metal-poor GCs outnumbering red, metal-rich GCs by a ratio
of 4:1.  The inner GCs associated with NGC~4874 also have a
bimodal distribution in color, but with a redder
metal-poor population. The fraction
of red IGCs (20\%), and the red color of those GCs, implies that IGCs
can originate from the halos of relatively massive, $L^\ast$ galaxies,
and not solely from the disruption of dwarf galaxies. 

\end{abstract}



\keywords{galaxies: elliptical and lenticular, cD ---
  galaxies: clusters: individual: Coma ---
  galaxies: halos ---
  galaxies: evolution --- galaxies: star clusters: general -- 
  globular clusters: general}

\section{Introduction}
\setcounter{footnote}{0}
\subsection{Intracluster Stellar Populations and Hierarchical Galaxy Formation}

Massive elliptical galaxies at the centers of galaxy clusters---often
brightest cluster galaxies (BCGs) and sometimes cD galaxies---generally have
little ongoing 
star formation with only minor evolution since $z\sim1$, and only a 
shallow relationship between their stellar mass and their host cluster
mass (e.g., Lin \& Mohr 2004; Whiley \etal 2008).  In the standard hierarchical
paradigm, however, the most massive halos should be the last to
assemble, and so these galaxies have traditionally presented problems
for formation models.  

Recent simulations suggest that this paradox can be resolved in a picture
where the stars that end up in the most massive galaxies form early,
energy feedback from supernovae and active galactic nuclei
subsequently suppress star formation, and the assembly of these
galaxies through dry mergers continues right up to the present day
(De~Lucia \& Blaizot 2007; although see Bildfell \etal 2008 for
evidence that feedback is not 100\% efficient).  Thus, even if there
is little star 
formation at late times, these galaxies are still expected to further
assemble stellar mass at $z<2$.  This predicted increase in the masses
of BCGs over time, however, may be in conflict with observations that show
little mass evolution of BCGs from $z\sim1.5$ to the present (Collins
\etal 2009).

It is expected that the dry merging or tidal stripping of satellites should
not only contribute stars to the central galaxy itself, but also to
an intracluster component that has previously been associated with
the extended stellar envelopes of cD galaxies (Matthews, Morgan \&
Schmidt 1964), and is sometimes labeled as a ``diffuse
stellar component'' (Monaco \etal 2006) or simply ``intracluster
light'' (ICL).  This component can make up a large fraction of the
total luminosity at the center of galaxy clusters (Oemler 1976), 
and if added to the stellar mass of central
cluster galaxies, might naturally explain current contradictions
between simulation and observation.  Purcell, Bullock,
\& Zentner (2007) simulated the formation of the ICL from the
shredding of satellite galaxies, finding that in massive clusters, the
ICL can dominate the total stellar mass of the combined ICL+BCG
system, which is consistent with observations of low redshift clusters
(Gonzalez, Zabludoff, \& Zaritsky 2005; Seigar, Graham, \& Jerjen 2007).

In fact, there is increasing observational
evidence that a significant fraction ($10$--$40\%$) of the total
stellar light in a galaxy cluster is intergalactic.  Starting with Zwicky
(1951), many detections of low surface brightness starlight in galaxy
clusters---both in cD envelopes and in the regions between
galaxies---support the existence of substantial intracluster stellar
populations (Welch \& Sastry 1971; Uson \etal 1991; 
Vilchez-Gomez \etal 1994; Gregg \& West 1998; Trentham \& Mobasher 1998;
Feldmeier \etal 2002, 2004a; Lin \& Mohr 2004; Adami \etal 2005;
Zibetti \etal 2005; Mihos \etal 2005; Gonzalez \etal 
2005; Seigar \etal 2007; Gonzalez, Zaritsky, \& Zabludoff
2007; Krick \& Bernstein 2007). 
In nearby clusters, there have also been direct detections of
intergalactic red giant 
branch stars (Ferguson, Tanvir, \& von Hippel 1998), asymptotic giant
branch stars (Durrell \etal 2002) planetary
nebulae (Theuns \& Warren 1997; M\'{e}ndez \etal 1997; Feldmeier,
Ciardullo, \& Jacoby 1998; Feldmeier \etal 2004b; Okamura \etal 2002;
Arnaboldi \etal 2004; Gerhard \etal 2007; Arnaboldi \etal 2007,
Castro-Rodrigu{\'e}z \etal 2009; Doherty \etal 2009), novae
(Neill, Shara, \& Oegerle 2005), and supernovae (Gal-Yam \etal 2003).
Detections of intergalactic light have also been made in compact
groups (Da Rocha \& Mendes de Oliveira 2005).
These studies tend to show that  the richer and more massive the
group or cluster, the larger the fraction of intergalactic light.

\subsection{Intergalactic Globular Clusters}
Another important clue to the formation of massive ellipticals is that
those residing at the centers of galaxy clusters often host extremely large
populations of globular clusters (GCs).  
The star forming events that form globular clusters will
mostly form stars that end up in the field, so it is natural that the
number of GCs in a galaxy should roughly scale with that galaxy's
stellar luminosity or mass.  However, the ratio of GCs to
starlight---usually characterized as the specific frequency,
$S_N$ (Harris \& van den Bergh 1981)---has long been known to vary
across galaxy mass and
morphology, with giant central elliptical galaxies
harboring the largest GC systems and having some of
the highest specific frequencies. This abundance of GCs in galaxy clusters
appears explainable if the number of GCs scales with either total
baryonic mass at the cluster center, including hot gas (McLaughlin
1999), or the total 
dynamical mass of the cluster (Blakeslee \etal 1997).
Blakeslee (1997, 1999)
observed that the number of GCs in galaxy clusters was directly
related to cluster mass, but the relatively constant BCG luminosity
thus led to high $S_N$.  It is possible that the high specific
frequencies are because measurements of the galaxy luminosity
typically do not include a substantial ICL component, and that the high $S_N$ in
central cluster galaxies would be more normal if the ICL was included.
Galaxy specific frequency also varies with galaxy stellar mass (or
luminosity) in a way that is consistent with the expected variation in 
galaxy stellar mass fraction (or mass-to-light ratio) (Peng \etal
2008; Spitler \etal 2008).  

This connection between GCs and total mass
has interesting implications, particularly in massive galaxy clusters
where the predicted build up of stellar mass in central galaxies
should be paralleled by the build up of a large GC system.
If much of the stellar mass in galaxy clusters resides in the low surface
brightness ICL then there should also be a corresponding population of
intracluster GCs (IGCs) that are not
gravitationally bound to individual galaxies, but directly to the
cluster itself.  Moreover, the detection of point
source IGCs in the nearest clusters is a much easier observational endeavor
than measuring the faint ICL, giving us a window onto the nature of the
diffuse stellar content.

There are other reasons to expect substantial populations of IGCs.
West \etal (1993) proposed that GC formation may be
biased toward the largest mass overdensities, i.e.\ galaxy clusters.
West \etal (1995) also proposed that populations of IGCs were
responsible for the high $S_N$ seen in cD galaxies.  More recently,
spectroscopy of ultra-compact dwarfs (UCDs) and massive GCs (also
dubbed dwarf-globular transition objects; \hasegan \etal 2005) have
uncovered a population of compact stellar systems in galaxy clusters
resembling the most massive GCs or dE nuclei stripped of their host
galaxies (Drinkwater \etal 2003; Hilker \etal 2007; Mieske \etal 2008;
Gregg \etal 2009; Madrid \etal 2010; Chiboucas \etal 2010).  These
objects, while generally more massive than 
typical GCs and consequently may have different origins, might
be the so-called  ``tip of the iceberg'' for a large population of
free-floating, normal globular clusters.

In fact, a number of extragalactic GC studies over the past few years
have strongly suggested the presence of IGCs in nearby galaxy
clusters.  In the Virgo and Fornax Clusters, serendipitous discoveries of GCs
in intergalactic regions using {\it HST} imaging (Williams \etal 2007),
ground-based imaging (Bassino \etal 2003) and spectroscopy (Bergond
\etal 2007) point to the existence 
of IGC populations.  However, it is often unclear whether these GCs
are truly intergalactic, or are part of the extended halos of cluster
galaxies (c.f.\ Schuberth \etal 2008).  A recent study by Lee, Park \&
Hwang (2010), however, used data from the Sloan Digital Sky Survey
(SDSS) and found statistically significant detections of
GC candidates throughout the Virgo cluster.  In more distant galaxy
clusters, candidate IGC populations have been identified as point source
excesses in HST imaging (\jordan \etal 2003; West \etal 2011).

It is possible that some IGCs and intracluster stars formed {\it in
  situ}, i.e.\ in cold, intergalactic gas that never accreted onto or
was stripped from galaxies.  It is also possible that IGCs formed very
early and at high efficiencies in dwarf-sized subhalos (e.g., Moore
\etal 2006; Peng \etal 2008), and whose host
galaxies were subsequently tidally destroyed by interactions with
the cluster potential.  Another possibility is that IGCs were formed
in larger galaxies and were stripped through tidal interactions with
the cluster potential or with other galaxies
(see, e.g., simulations of Yahagi \& Bekki 2005; Bekki \& Yahagi 2006).

The formation of the IGC population is obviously linked to that of the
ICL, although the observed properties of the two populations may be different.
For example, the detectability of the ICL is highly dependent
on its surface brightness, whereas IGCs are detectable even in
isolation.  Simulations by Rudick \etal (2009) show that the ICL is
supplied by tidal streams that originally have relatively high surface
brightness but then disperse to become fainter and harder to detect.
ICL studies using surface photometry are thus more
sensitive to recent disruptions, whereas the IGC population is a less
temporally biased tracer of the full intracluster stellar population.

The study of extragalactic GC systems has been transformed by the high
spatial resolution imaging of the {\it Hubble Space Telescope} ({\it HST}), with
observations of hundreds of GC systems now in the archives and
published literature (e.g., \jordan \etal 2009).  {\it HST}'s
deep sensitivity to compact or unresolved sources, and its ability to
distinguish background galaxies from likely GCs, makes it an ideal
tool for extragalactic GC studies. However, the relatively
narrow field of view of {\it HST}'s cameras and the close proximity of
the galaxies being studied 
($D\lesssim100$~Mpc) means that most {\it HST} studies have focused on GC
systems directly associated with galaxies.  Observations of wider
fields are usually conducted with ground-based telescopes (e.g.\
McLaughlin 1999; Bassino \etal 2006; Rhode \etal 2007), that
gain area at the cost of spatial resolution.

\subsection{The Coma Cluster of Galaxies}

The Coma Cluster of galaxies (Abell 1656) is one of the nearest rich,
dense clusters, and is a fundamental target for extragalactic
studies.  Studies of GC systems in Coma from
the ground using surface brightness fluctuations (Blakeslee \etal
1997; Blakeslee 1999) and using HST (Kavelaars \etal 2000; Harris
\etal 2000; Harris \etal 2009) all point to large GC systems around
the cluster's giant elliptical galaxies, particularly around the
central massive galaxy, NGC~4874 (Harris \etal 2009).
Although many photometric studies support the existence of an intracluster
stellar light component in Coma (e.g., Zwicky 1951; de Vaucouleurs \&  de
Vaucouleurs 1970; Welch \& Sastry 1971; Kormendy \& Bahcall 1974;
Matilla 1977; Melnick, White \& Hoessel 1977; Thuan \& Kormendy 1977; 
Gregg \& West 1998; Calc{\'a}neo-Rold{\'a}n \etal 2000; Adami \etal 2005), and 
even velocities for intracluster PNe have been measured (Gerhard \etal
2007; Arnaboldi \etal 2007), evidence for or against the existence of IGCs
is much more muddled. A search for IGCs in Coma using ground-based
data by Mar\'{i}n-Franch \etal (2002, 2003) did not find a surface
brightness fluctuation signal that would have hinted at the presence
of IGCs, mainly because of the shallow limiting magnitude of their
photometry.

\begin{figure*}
\plottwo{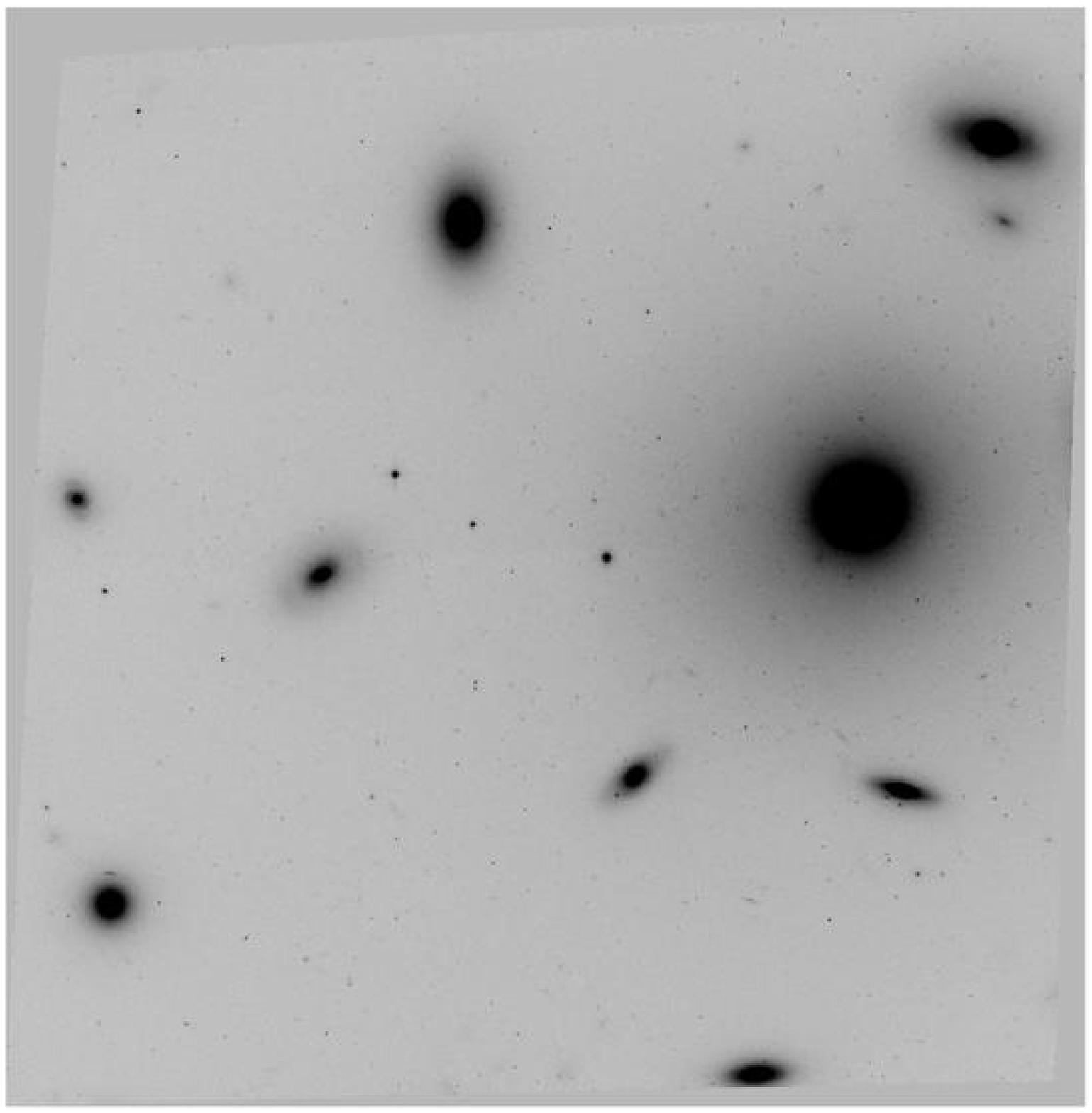}{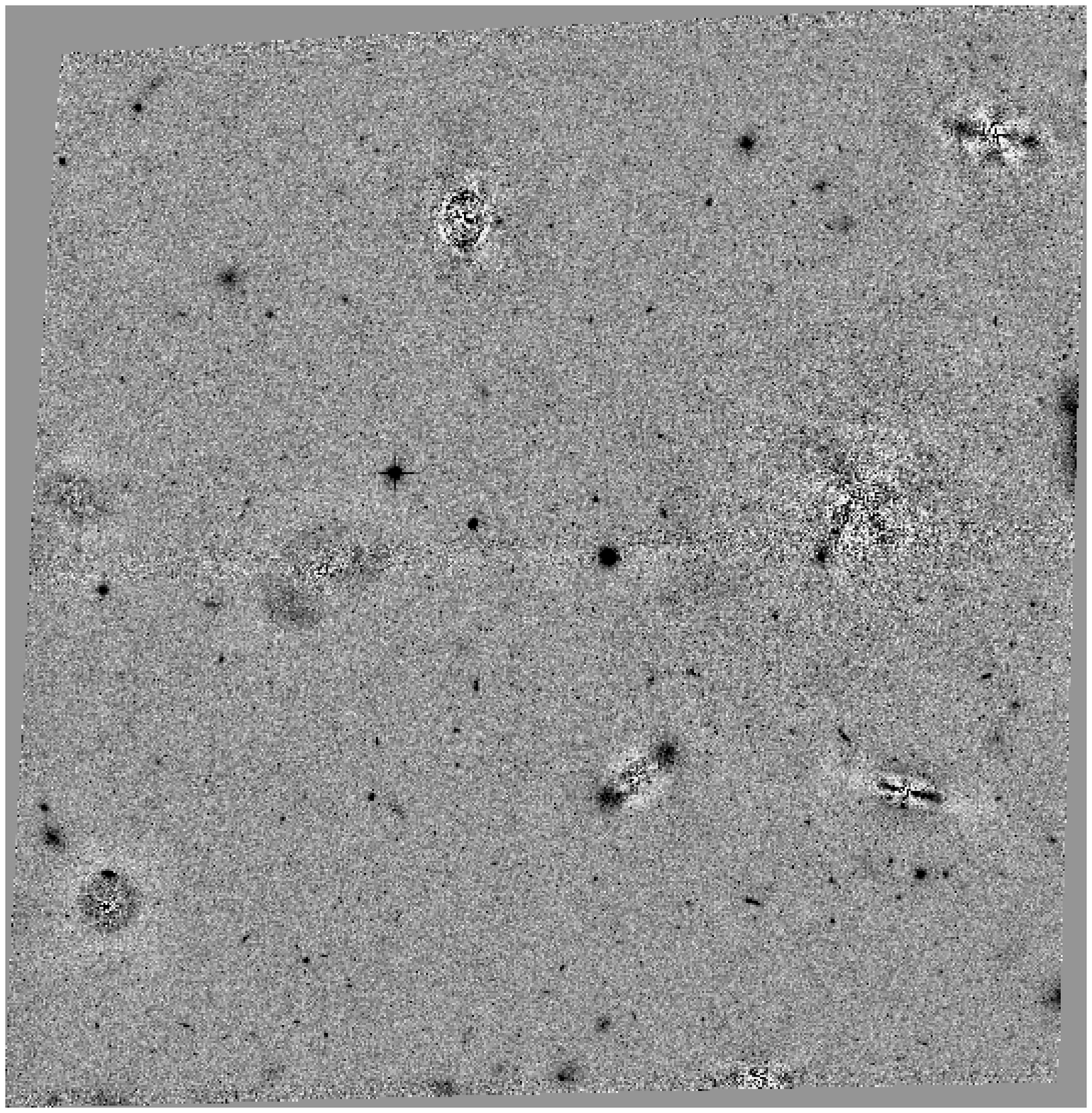}
\caption{(a) The ACS F814W image of the central pointing (Visit 19) containing
  NGC~4874 and many other bright elliptical galaxies. North is left,
  and East is down.  The image is $202\arcsec$ (98~kpc) on a side.  (b) The same
  pointing but after our iterative galaxy subtraction.  
  Residuals at galaxy centers are still visible at this contrast
  level, but the overall large scale gradients in the background
  light have been removed. \label{fig:v19}}
\end{figure*}

At a distance of 100~Mpc, the value we adopt for this paper ($m-M=35$,
Carter \etal 2008), $1\arcmin$ on the sky subtends
29~kpc in the cluster, and the mean of
the GC luminosity function (GCLF) in giant ellipticals is
$I_{Vega}=26.44$~mag.  This regime of projected areal coverage and GC
apparent brightness makes it reasonable to conduct a contiguous 
survey of the Coma cluster core for GCs, unbiased by the locations of
individual galaxies, using the {\it HST} Advanced Camera for Surveys
(ACS) Wide Field Channel (WFC).

The {\it HST/ACS} Coma Cluster Survey is a
Treasury Survey originally approved for 164 orbits.  One of the main
components of this survey was a contiguous ACS/WFC mosaic of the core
of the Coma cluster, making it the ideal data set to investigate the
existence of intergalactic GCs.  Hints of this population have already
appeared in studies of UCDs by Madrid
\etal (2010) and Chiboucas \etal (2010).  This paper presents the first
compilation and description of IGCs in the Coma cluster.

\section{Observations and Data}

\subsection{Imaging Data}
The data used in this study are from the {\it HST/ACS} Coma Cluster
Survey.  The survey observations and data reduction are described in
detail by Carter \etal (2008), the catalog generation for the
public data release is described by Hammer \etal (2010), and an
in depth analysis of galaxy structural parameters and completeness is
presented in Balcells \etal (2010).  We summarize the relevant information here. 

The Coma Cluster Survey, as originally designed, consisted of a large central
ACS mosaic of the Coma cluster core, and 40 targeted observations in
the outer regions of the cluster.  The central mosaic was designed to
be 42 contiguous ACS/WFC pointings in a $7\times6$ tiling configuration,
and covering an $21\arcmin\times 18\arcmin$ area.  Each pointing is
observed in two filters, F475W ($g$) and F814W ($I$), with exposure
times of 2560s and 1400s, respectively.  Unfortunately, the
failure of the ACS/WFC (January, 2007) meant that only 28\% of the
survey was completed: 
19 pointings in or around the central mosaic, and 6 in the outer
regions.  Within the 
core, the central galaxy NGC~4874 was observed, but the other giant
elliptical, NGC~4889, was not imaged before the ACS failure.  Despite the
shortfall, the current observations still provide the largest set of
deep, high resolution imaging available for this important galaxy cluster.
Recent studies of compact galaxies in Coma (Price \etal 2009) and
spectroscopy of Coma cluster members (Smith \etal 2009) are part of
a concerted effort to study galaxy evolution in the Coma cluster built
around this $HST$ Treasury survey.

The ACS data reduction was performed using a dedicated Pyraf/STSDAS
pipeline that registered and combined images while performing
cosmic-ray rejection.  The dithered images were combined using
Multidrizzle (Koekemoer \etal 2003), which uses the Drizzle
algorithm (Fruchter \& Hook 2002). For this study, we used the data
drizzled for the ACS Coma Survey Data 
Release 2 (DR2). However, except for Visits 3, 10, and 57, the F814W
images on which the bulk of this paper is based are identical to those
in Data Release 1 (DR1). 

\subsection{Object Catalogs and Galaxy Subtraction}
Our images of the Coma cluster reveal a striking amount of
detail: cluster members across the mass spectrum, globular clusters,
background galaxies, and a few foreground stars.  Given the different
spatial scales of these objects on the sky, it is important to
generate catalogs with detection parameters optimized for the
subject under study.  For our purpose of
studying the globular clusters {\it between} galaxies, we used the
well-tested Source Extractor software package (Bertin \& Arnouts
1996) with parameters optimized for point source detection, and which
are effectively identical to those used by Hammer \etal (2010) in the
public data release\footnote{For details, visit the Coma Cluster
  Survey website at {\tt http://astronomy.swin.edu.au/coma/}}.  
Photometry was put on the AB magnitude system 
using the zeropoints of Siranni \etal (2005).  All magnitudes in this
paper are AB unless otherwise specified.

For most visits, the area between
galaxies is much larger than that occupied by galaxies and the catalogs
can be considered effectively complete to the same level except in the
close vicinity of cluster galaxies.  The one exception is Visit 19,
which contains NGC~4874 and many other ellipticals in
the cluster core (Figure~\ref{fig:v19}a).  This pointing is nearly 
entirely dominated by the light from one galaxy or another; it is also the one
with the highest concentration of GCs.  To address this, we
implemented an iterative galaxy subtraction algorithm that produced a
fully background subtracted image (Figure~\ref{fig:v19}b).

We subtracted the 10 brightest galaxies from Visit 19.  We started from the
brightest (NGC~4874 itself) and worked to the faintest of the ten.  In
each case, we first manually masked all the bright galaxies except for
the one being subtracted.  We then used the IRAF {\tt ellipse} and
{\tt bmodel} tasks (Jedrzejewski 1987) to model the isophotes of the
object galaxy.  Because the {\tt ellipse} fitting only occurs out to a
finite radius, the resulting model will have finite extent, and the
subsequent subtraction will leave a sharp discontinuity in the image.
For convenience of object detection, we extended the 
{\tt ellipse}-generated models with a power-law fit
to the last five data points in the profile at fixed ellipticity.
This allowed for a
smooth subtraction out to the borders of the ACS image.  After
subtracting one galaxy, we then repeated the process with the next
brightest galaxy on the subtracted image.  After the last galaxy was
subtracted, we used Source Extractor to create and subtract a
background map that removed large scale variations.  This last step is
important because it allows us to recover from any large scale over-
or under-subtractions due to mismatches between the power-law
extensions and the true surface brightness profiles of the galaxy.
A similar technique was used with success by \jordan \etal (2004)
in ACS images of Virgo cluster galaxies, although that was only for
single galaxies.

After galaxy subtraction, we generated catalogs with Source Extractor,
using variance maps that accounted for the extra Poisson noise
expected from the subtracted galaxy light.  These catalogs contain
objects much closer to the centers of galaxies, and to NGC~4874 in
particular.  Photometry was obtained by using 3 pixel radius
($0\farcs15$) circular apertures with aperture corrections and
zeropoints from Sirianni \etal (2005).  Unless otherwise specified,
all magnitudes in this paper are on the AB system.  These objects are
included in the DR2 catalogs of Hammer \etal (2010).

\subsection{Completeness}

We use artificial star tests to quantify the spatially varying
detection efficiency across our images.  This is particularly
important for the galaxy-subtracted image containing NGC~4874, where
the bright galaxy light affects the depth of our observations.  

We first use routines in DAOPHOT II (Stetson 1987) to construct an
empirical PSF using bright point sources in Visit 19.  At the distance
of the Coma Cluster, nearly all globular clusters are unresolved with
{\it HST}, and
can be well-approximated by point sources (the mean half-light radius
of GCs, $r_h\approx 3$~pc, is only $\sim6\%$ the full width half
maximum of the point spread function). Because detection is done 
only in the F814W band, we only add artificial stars to these images.  

When adding point sources into the images, we avoid objects in the
image as well as artificial stars already placed so as to
avoid incompleteness due to confusion.  We run the exact same
detection pipeline on these images as we do to create our object
catalog and record whether the objects were detected as a
function of magnitude and position.  The number of artificial stars
added and measured---7,000,000 in Visit 19 alone, and approximately
4240~arcmin$^{-2}$ for the other visits---ensures that we can derive a
completeness curve for any position in the survey, and for any GC
selection criteria.

In a typical blank area in our images observed with the full
exposure time, the 90\% completeness level is at $I\approx26.8$~mag, and
the 50\% completeness level is at $I\approx27.3$~mag.  At
$R\approx2\arcmin$ from the center of NGC~4874, however, these limits are 
1.5~mag shallower.

\subsection{Globular Cluster Candidate Selection}

One of the main benefits of GC studies with HST is the ability to use
morphology and resolution to separate GCs from their main
contaminants, background galaxies.  At Coma distances, GCs are
point sources when observed with {\it HST}, but the great majority of
background galaxies are resolved.  We use this ability to select
against background
contaminants and produce a relatively clean sample of GC candidates.

\begin{figure}
\plotone{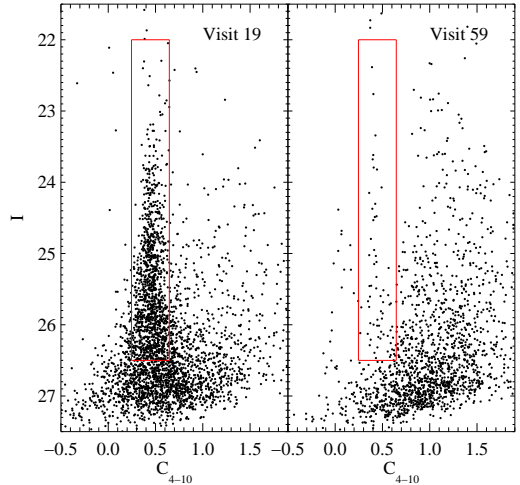}
\caption{$I$ magnitude versus concentration index
  ($C_{4-10}= m_{4pix}-m_{10pix}$) for Visit 19, the central pointing containing
  NGC~4874 (left) and Visit 59, the background pointing most
  distant from the cluster center which contains mostly
  background galaxies (right). The vertical locus of points around
  $C_{4-10}=0.45$ contains point sources.  Most of the point sources
  in Visit 19 are likely to be GCs.  The background
  galaxies are mostly resolved to be more extended than the GCs until
  $I\sim27$, where some overlap the stellar locus.  The red outlines
  shows our selection region for GCs.
 \label{fig:gcselect}}
\end{figure}

We use a rough but effective concentration criterion to select GCs.
Figure~\ref{fig:gcselect} shows the ``magnitude-concentration'' diagram for
objects, where we measure a concentration index, $C_{4-10}$ using the
difference in magnitude measured in a 4 pixel diameter aperture and a
10 pixel diameter aperture.  Figure~\ref{fig:gcselect} shows that this
index works well to distinguish point sources from extended
sources.  Here, we show the distribution of objects in
Visit~19 (the one containing NGC~4874), which has the
largest number of GC candidates.  We overplot the objects from
Visit 59, which is the most remote of our fields, and contains mostly
background galaxies.  The red lines show our selection region where we
exclude nearly all of the background galaxies.  Although we
experimented with different cuts in this diagram, including a variable
width of the selection region with magnitude, the variations were not
significant and we decided in the end that simplicity was best,
choosing a cut of $\pm0.2$~mag around the median concentration for
point sources ($\langle C_{4-10}\rangle =0.45$).

For the purposes of this study, we wish to maximize the number of
good GC candidates, while also balancing the increasing number of
background contaminants with magnitude.  Because of the depth and
high spatial resolution of our data, we chose a fairly conservative
magnitude limit, including objects with $I<26.5$~mag.  At this magnitude,
our data is 97\%
complete in regions free of galaxy light, so completeness corrections
are only important toward the centers of galaxies.  At the Coma
Cluster distance ($m-M=35$), assuming an extinction $A_I=0.017$~mag
for NGC~4874 (Schlegel, Finkbeiner, \& Davis 1998), this limit should include
a significant fraction
of the GCs in a Gaussian GC luminosity function
(GCLF) typical of giant ellipticals.  We use the recently measured
$I$-band GCLF measurement for the Virgo cD galaxy M87 (Peng \etal
2009), which was performed with deep HST/ACS observations in the same
F814W filter used by the ACS Coma Survey.  Peng \etal (2009) quote a
GCLF Gaussian mean and sigma of $\mu_{I,Vega}=-8.56$~mag and
$\sigma=1.37$~mag.  For AB magnitudes, we add 0.436~mag to
$\mu_{I,Vega}$ (Sirianni \etal 2005).
Assuming these values for a Gaussian GCLF, our GC catalog magnitude
limit should include $\sim39\%$ of all GCs, and $\sim75\%$ of the
luminosity in GCs.

\begin{figure}
\plotone{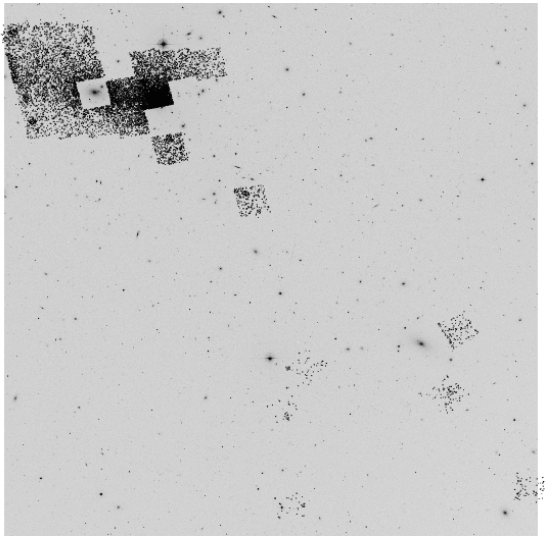}
\caption{Spatial distribution of ACS GC candidates shown on a
  $1^\circ\times 1^\circ$ 
  Digitized Sky Survey image of the Coma Cluster with North up and
  East to the left ($1.75\times1.75$~Mpc at Coma distance).  At the top left
  of the image is the observed portions of the cluster core central
  mosaic. The largest concentration of GCs is around the central galaxy,
  NGC~4874.  The other large galaxy, unobserved by ACS, is NGC~4889.
  At the bottom of the image are six fields in the outer regions
  of the cluster.  The three outer fields to the right (west) show higher
  numbers of GC candidates because of their proximity to large
  galaxies.  The eastern three outer fields (bottom center) are not
  near large galaxies and are used
  as background fields.  The density of GC
  candidates throughout the entire cluster core is much higher than in the
  background regions, implying a large
  population of intracluster GCs. The ACS field sizes are roughly
  202\arcsec on a side.  \label{fig:spatialdss}}
\end{figure}

This is likely an oversimplification, however, as both the mean and
width of the GCLF 
is known to vary with galaxy mass (\jordan \etal 2006; 2007).  
If we assume a Gaussian GCLF typical of dwarf ellipticals in clusters
($\mu_{I,Vega}=-8.1$~mag and $\sigma=1.1$~mag, Miller \& Lotz 2007),
then our limit 
includes $\sim22\%$ of the total number.  This discrepancy is one of
the main systematic uncertainties in our analysis.  We emphasize,
however, that changing the assumed GCLF does not affect the
significance of our result, just the inferred total number of GCs.
Given that the depth of our
data is not sufficient to measure the GCLF parameters directly,
we choose to assume the brighter 
GCLF, seen in giant ellipticals, as this will give us a lower estimate for
the number of GCs in any given area. The numbers could be higher by
$\sim80\%$ in regions where the GCLF for dwarf ellipticals is more
representative. 

We also introduce a broad color cut of $0.6 < (g-I) < 1.5$ that should
include all old globular clusters.  This color range is based on the
transformed $g$--$z$ colors of GCs in the ACS Virgo Cluster Survey
(\cote \etal 2004; Peng \etal 2006), and mainly eliminates distant,
compact red galaxies.  The ages of extragalactic GCs across all
metallicities are primarily old ($>5$~Gyr), especially those
associated with massive early-type galaxies (e.g.\ Peng \etal 2004;
Puzia \etal 2005; Beasley \etal 2008; Woodley \etal 2010), so this
color range should include all bona fide GCs.

\section{Spatial Distribution of GC Candidates}
Figure~\ref{fig:spatialdss} plots the locations of GC candidates in our ACS
images on a DSS image of the cluster.
While it is not surprising that the number of GCs is high around massive
ellipticals such as NGC~4874, what is striking about this figure is that
the number of GCs across the {\it entire} central mosaic is high and is
significantly elevated when compared to the numbers in the outer
fields.  Even the corner fields of the central mosaic have many more
GCs.  Of the six outer fields, three in the southwest (lower right)
have visibly elevated GC numbers
due to their proximity to NGC~4839 (top) and NGC~4827, two giant early-type
galaxies.  The three other fields to the south are not near massive
cluster members.  We take these three southern fields as an upper
limit on the background contamination from foreground stars and
compact distant galaxies.  All of these fields have fewer GC
candidates than does any field in the central mosaic.

Other than an obvious concentration around
NGC~4874 and NGC~4889 (the latter of which was not observed with ACS),
the GC distribution is relatively uniform across most of the
central mosaic and not spatially clustered; i.e., with the exception of the
two central ellipticals, the spatial structure of
the GCs is not highly correlated with the positions of cluster
galaxies.  This is partly a bias introduced by the failure of
Source Extractor to detect GCs that are immediately in the vicinity of
bright galaxies.  However, GC detection should not be a problem in the
halos of the galaxies, and except in a few cases we do not detect the
kind of small-scale substructure one would expect in the cluster GC
distribution if all the GCs were tightly associated with galaxies.

\begin{figure}
\plotone{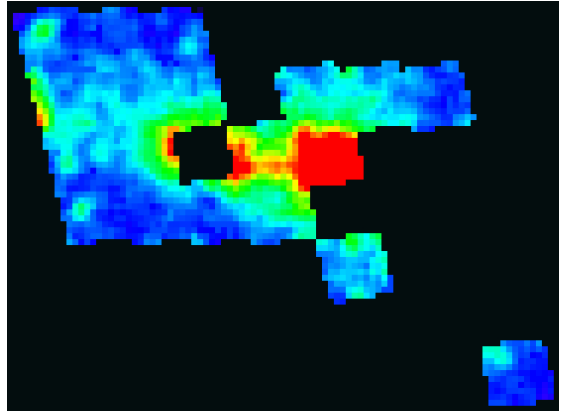}
\caption{
Smoothed spatial distribution of GCs in the Coma Cluster core
($30\farcm8\times23\farcm0$, $900\times670$~kpc).  Pixels
are $20\arcsec$ on a side, and color represents the surface density
of GCs, corrected for completeness (blue to red denotes low to high
density).  The entire image has been 
smoothed by a Gaussian kernel with $\sigma=30\arcsec$.  The dominant
concentration of GCs is around NGC~4874, and an extended structure of
GCs appears to connect NGC~4874, 4889, and 4908.  Some
peaks in the distribution represent individual cluster galaxies.
\label{fig:gcgrid}
}
\end{figure}

Figure~\ref{fig:gcgrid} shows more clearly the distribution of GCs in
the cluster core.  To produce this figure, we divide the core region into
$20\arcsec\times 20\arcsec$ ``pixels'' with each representing the
surface density of GCs, corrected for spatially varying completeness,
and smoothed with a Gaussian kernel with $\sigma=30\arcsec$.
The large concentration of GCs at the center-right is the GC system of
NGC~4874.  The GC system of NGC~4889 is also evident, although the
galaxy itself was not observed.
While Figure~\ref{fig:spatialdss} shows that the overall surface density of
GCs is well above the background, Figure~\ref{fig:gcgrid} shows
hints of large-scale substructure in the GC spatial distribution.
There appears to be an extended structure of IGCs connecting
NGC~4874 to NGC~4889 to NGC~4908 and IC 4051, both of which lie just
beyond the eastern edge of the mosaic.

These observations suggest the existence of a large intergalactic
population of globular clusters.  In the following sections, we seek
to verify and quantify their existence.

\section{Background Estimation and Galaxy Masking}
\label{sec:bgmask}

Contaminants to our sample of GCs consists of foreground stars and
faint, unresolved background galaxies (the sum of 
which we generically refer to as ``background'').  This
background is important to quantify, as a smooth background can mimic
a smooth IGC population.  Ground-based IGC studies in Coma are
typically plagued by high background due to their inability to
distinguish distant galaxies from point sources.  

As a measure of our background, we choose the three outer ACS 
fields---visits 45, 46, and 59---that are not near giant galaxies, and
are shown at the bottom-center of Figure~\ref{fig:spatialdss}.  For
each of these fields, we select GC candidates as described earlier,
and also mask the regions containing a few obvious Coma members using
the prescriptions described below.  The
surface density of GC candidates over these three fields is 
$2.8\pm0.3$~arcmin$^{-2}$, or $\sim28$ per ACS field.  As we will
show, this is nearly an order of magnitude lower than the density of
GCs even in the outer fields of the Coma core.

To verify this background level, we compared our point source
counts in these fields to those in the COSMOS HST Treasury project
(Scoville \etal 2007).
The COSMOS survey imaged 1.8~deg$^2$ at high Galactic latitude with
the same camera (ACS/WFC), filter (F814W), and depth as the Coma
Cluster Survey.  The number of point sources in our three background
fields is entirely consistent with the numbers expected from the 
surface density of stars in the COSMOS fields. Down to
$I_{814}<25$~mag, we detect $104\pm10$ point sources in our
three background fields and 98 are expected using the average surface
density from the COSMOS data.  This independent check gives us more
confidence that our background value is correct. 

The fact that the global background is so low compared to the
detections in our Coma core fields gives us confidence that
we are indeed detecting GCs within the Coma cluster.  The more
difficult question is whether these objects are truly
``intergalactic'' or simply part of extended galactic systems.
This debate is not one easily resolved by imaging data
alone.  We can, however, address the contribution from galactic GC
systems in two ways.  First, we aggressively mask regions around known
bright galaxies.  Second, we can make certain assumptions about the numbers and
spatial extent of the GC systems of observed cluster members and
compare simulated GC distributions to the observations.  We do this in
order to test the hypothesis that the GCs observed in the cluster core
are an intergalactic population.

The details of these two methods are described in
Appendix~\ref{sec:appendix}.  In short, we generate masks around all
galaxies with luminosities down to $M_g<-17$~mag, both in and around all
of our fields.  The detection algorithm that we use for GCs actually
ends up masking GCs around fainter galaxies because our chosen
background estimation parameters cannot follow the steeply rising
surface brightness profiles at the centers of galaxies.
For each galaxy, we apply a liberal, size-dependent
mask to the surrounding regions.  These masks should eliminate $\sim90\%$
of the ``galactic'' GCs from our catalogs.  For the remaining outer
GCs, we subtracted a model GC system using an assumed \sersic\ $n=2$ profile
for the GC surface density, a reasonable assumption given previous
measurements of GC system radial profiles.  The parameters of this
model are estimated based on scaling relations for $S_N$ and $R_e$ from
Peng \etal (2008, 2011).  This modeling is done for all Coma galaxies
in the Eisenhardt \etal (2007) catalog, which is complete to $M_V<-16$~mag
and extends to $M_V<-14$~mag.  This is described in greater detail
in Appendix~\ref{sec:simgc}.

Although we have taken great pains to model and subtract any residual GCs that
may belong to Coma galaxies, we find that {\it our final result is
 largely insensitive to the assumed parameters}.  The detection of
IGCs, as we will show below, is highly significant and not dependent
on the details of the background or the modeling of GC systems.  We
estimate the systematic uncertainty due to our modeling procedure to
be $^{+4000}_{-5000}$ GCs (Appendix~\ref{sec:systematics}), only
5--9\% of the inferred IGC population.

\section{Results}

\subsection{Radial Profile of GCs in the Coma Cluster}
\label{sec:gcprofile}

NGC~4874 has previously been observed to have a large
number of globular clusters (Blakeslee \& Tonry 1995; Harris \etal
2000, 2009).  It has also been shown to have a GC system whose spatial
profile is shallower and more extended than those for other elliptical
galaxies (Harris \etal 2009).  Could the GCs that we see filling the
cluster core simply be the extended GC system of NGC~4874?  The
situation is complicated by the fact that the core of the cluster
contains not one but two giant ellipticals, the other being NGC~4889,
as well as many other member galaxies.


\begin{figure}
\epsscale{1.1}
\plotone{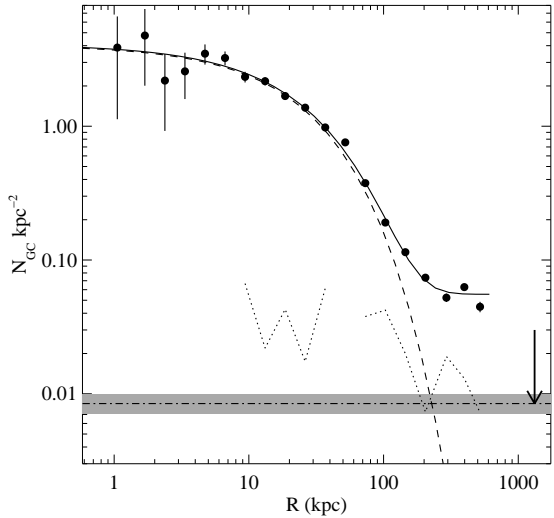}
\caption{\footnotesize The radial distribution of GCs in the Coma Cluster core
  centered on NGC~4874.  The surface density of GCs in each bin (black
  points) is calculated after masking around known galaxies, and
  statistical subtraction of GCs belonging to these cluster members.
  The radial profile exhibits a flat inner core as well as an
  inflection and flattening at large radii.  We interpret the flat
  distribution at large radii as evidence of a large population of IGCs.
  The dot-dashed line and gray band at bottom denote the surface
  density of background 
  objects (plus $1\sigma$ errors) determined from our outer ACS fields
  and subtracted from all radial bins.  The background level is a
  factor $\sim7$ below that in the outermost bins.  The arrow at
  bottom right 
  shows the mean distance from NGC~4874 of the three background fields.
  The dotted
  line is the modeled radial distribution of GCs belonging to cluster
  members that are still visible after masking.  These have been
  subtracted from the GC radial profile, although they too are
  well below the overall level by a factor of a few.  The data are
  well fit by a \sersic\ model plus a constant level (solid line).
  The \sersic\ component alone is shown as the dashed line.
\label{fig:radial}}
\end{figure}

In Figure~\ref{fig:radial}, we show the radial distribution of GCs in
the cluster core, centered on NGC~4874.  For each bin in radius, we
sum up the number of observed GC candidates in unmasked regions,
subtract the expected contribution of GCs from other Coma galaxies
(shown as the dotted line in 
Figure~\ref{fig:radial}), and subtract the global background level
(dot-dash line), leaving what should be the NGC~4874 and
IGC population.  We determine the mean completeness of
the sample within the annulus, and extrapolate the total number of GCs
assuming the M87 GCLF as described above.  We then sum the total
observed, unmasked area within the annulus to determine the surface
number density of GCs.  The random errors in each bin are derived from the
Poisson errors for the number of candidate GCs as well as from the
Poisson error in the background, added in quadrature.

This profile, tabulated in Table~\ref{tab:gcprofile}, represents our
best estimate of the radial surface 
density distribution of the GC system surrounding 
NGC~4874, uncontaminated by the GC systems of other cluster members.
Perhaps the most interesting feature of this profile is a marked inflection at
$R\sim200$~kpc, beyond which the
GC surface number density decreases much more slowly with radius.  We
interpret this flattening of the profile as the region where a large
and extended population of IGCs starts to dominate the GCs
directly associated with NGC~4874.  The significance of this detection
is extremely high, as the background level is shown in
Figure~\ref{fig:radial} by the horizontal dot-dashed line at the bottom, with
the estimated error of the background denoted as the shaded gray
region.  The inferred IGC surface density is a factor $\sim7$ over
the background. Another
point of comparison is with the modeled surface density of remaining
unmasked galactic GCs, shown as the dotted line, which also has already been
subtracted from 
our total GC profile.  The overall surface density of GCs in this
cluster profile is well above the surface density of masked galactic
GCs (by a factor 4--7), and thus the GCs we see are likely to
be truly intergalactic.  
We have also found
that our results do not change significantly if we only use data from
the eastern or western half of the Coma core.

\begin{deluxetable}{ccc}
\tablewidth{0pt}
\tablecaption{GC radial surface density profile centered on NGC 4874}
\tablehead{
\colhead{$\langle R \rangle$} & \colhead{$\sigma$} & \colhead{Area} \\
\colhead{(arcmin)} & \colhead{(arcmin$^{-2}$)} & \colhead{(arcmin$^2$)}
}
\startdata
  0.036 & $  3276\pm  2321$ &   0.00546 \\
  0.058 & $  4031\pm  2331$ &   0.00719 \\
  0.082 & $  1856\pm  1076$ &   0.01428 \\
  0.115 & $  2175\pm   825$ &   0.02834 \\
  0.162 & $  2950\pm   507$ &   0.05617 \\
  0.229 & $  2737\pm   326$ &   0.11152 \\
  0.322 & $  1978\pm   181$ &   0.22114 \\
  0.454 & $  1837\pm   128$ &   0.38152 \\
  0.639 & $  1423\pm    91$ &   0.56911 \\
  0.900 & $  1164\pm    71$ &   0.73770 \\
  1.268 & $   828\pm    52$ &   0.96299 \\
  1.786 & $   641\pm    39$ &   1.45523 \\
  2.516 & $   318\pm    15$ &   4.62493 \\
  3.543 & $   161\pm     7$ &  12.59246 \\
  4.991 & $    97\pm     4$ &  22.04740 \\
  7.030 & $    62\pm     4$ &  16.82656 \\
 10.066 & $    44\pm     3$ &  19.62118 \\
 13.674 & $    53\pm     2$ &  40.98172 \\
 17.886 & $    38\pm     3$ &  14.17776 \\
\tablecomments{These surface densities are corrected for completeness,the
 full Gaussian GCLF, and include the masking and subtracting of GCs belonging
to other cluster members, as described in Section~\ref{sec:gcprofile}.
The surface density of contaminants (also correcting for the GCLF) as marked
by the dot-dashed line in Figure~\ref{fig:radial} is $7.2\pm1.2$.}
\enddata
\label{tab:gcprofile}
\end{deluxetable}

A single \sersic\ profile, normally a good fit to the surface density
profiles of GC systems, is not sufficient to describe the data for the
central Coma Cluster GCs.  Instead, we fit a model combining a
\sersic\ profile and a constant.  It is likely that the IGCs
have a radially decreasing density profile (although the
simulations of Bekki \& Yahagi (2006) suggest that they can also have
a flat density distribution within the central few hundred kiloparsecs), but
the data only allow us to measure their mean surface density.  The
solid line in Figure~\ref{fig:radial} traces the best 
fit model, and the dashed line that follows it until large radii is
the best fit \sersic\ component.  
The fitted surface density of IGCs is $0.055\pm0.002$~kpc$^{-2}$,
which is a $19\sigma$ detection over the background,
$0.00845\pm0.001$~kpc$^{-2}$, assuming Poisson
random errors.  As we discuss in Section~\ref{sec:bgmask} and
Appendix~\ref{sec:systematics}, we also need to account for systematic
uncertainties from our modeling, but they do not affect the main
conclusion, which is high
significance of the detection. We list the best fit parameters in
Table~\ref{tab:fit}. 

Although we cannot determine the shape of the IGC component's density
profile, one constraint is that it must fall rapidly after the limits
of our data.  The mean distance of the three fields we are using to
measure the background is shown as the vertical arrow at 1.3~Mpc.
Therefore, the GC surface density must fall to zero, or at least the level of
the dashed line, by this distance.  A steep falloff like this favors a
low $n$ \sersic\ profile for the IGCs ($n=1$--2), similar to the ICL
profiles in Seigar \etal (2007) and X-ray gas in galaxy clusters
(Demarco \etal 2003), but lower (i.e., steeper in the outer regions)
than dark matter halo density profiles (Merritt \etal 2006).
However, at these radii, it may not
make as much sense to speak of a circularly symmetric GC radial
profile, and it would be more useful to map in two dimensions the
spatial distribution of GCs. 

\subsection{Total Numbers of GCs and Specific Frequency}

We use our radial spatial density profile to estimate the total
number of GCs in the cluster core, which we define to be
the extent of our data. Integrating this profile for $R<520$~kpc
gives a remarkable $70000\pm1300$~GCs, with the IGC component
dominating the GC population beyond 150~kpc.    As 
listed in Table~\ref{tab:fit}, the number of GCs 
belonging to NGC~4874's ``\sersic\ component'' out to this radius is
$\approx23000$, leaving a remaining 
$47000\pm1600$ (random) $^{+4000}_{-5000}$
(systematic) to be IGCs.  There are over {\it twice} as many IGCs 
as there are GCs from the \sersic\ component, resulting in an
IGC fraction of the entire central GC system of $\sim70\%$.  

With a measurement of NGC~4874's luminosity, we can
calculate an ``intrinsic'' specific frequency for the galaxy.  Harris
\etal (2009) use a luminosity of $M_V=-23.46$ (adjusted to
$D=100$~Mpc), but surface brightness profiles from SDSS imaging
(J. Lucey, private communication) and KPNO 4-meter CCD imaging
(R. Marzke, private communication) show the galaxy to be substantially
brighter.  Measurements of the total $r$-band luminosity from
mosaicked SDSS frames gives a value of $r=10.23$ ($M_r=-24.77$,
assuming $E(B-V)=0.009$ from Schlegel \etal 1998), which
includes a 0.32~mag extrapolation using the best-fit \sersic\ profile
for the light beyond 
$R=7\arcmin$. The mean color of the galaxy is $g-r\approx 0.8$~mag, which
produces $M_V=-24.47$ using the Lupton (2005) transformation.

\begin{table}
\begin{center}
\caption{Best parameters for \sersic\ plus constant ($\Sigma_{IGC}$)
  model fit to Coma central GC system\label{tab:fit}}
\begin{tabular}{|l|c|l|}
\hline Parameter & Value & Description\\
\hline
$n$ & $1.3\pm0.1$  & \sersic\ index \\
$R_e$ & $62\pm 2$~kpc & \sersic\ effective radius \\
$\Sigma_e$ & $0.437\pm0.034$~kpc$^{-2}$ & GC surface density at $R_e$\\
$\Sigma_{IGC}$ & $0.055\pm0.002$~kpc$^{-2}$ & Mean IGC surface density\\
$N_{GC,tot}$   & $70000\pm1300$  & Total GCs within 520~kpc\\
$N_{GC,Sersic}$   & $23000\pm700$ & ``\sersic'' GCs within 520~kpc \\
$N_{IGC}$   & $47000\pm1600$ (r) $^{+4000}_{-5000}$ (s) & IGCs within 520~kpc \\
\hline
\end{tabular}
\end{center}
\end{table}

With this, we calculate
an ``intrinsic'' specific frequency for NGC~4874 of $S_N=3.7\pm0.1$
(the errors are purely from the total numbers of GCs and do not
include errors in the luminosity).
\footnote{If we use the older,
  fainter value for the luminosity, then $S_N=9.5\pm0.3$, which is
  consistent with the value found in the
  HST/WFPC2 study of Harris \etal (2009).  Their data 
  only extended to $R\sim65$~kpc, and thus were not able to detect the
  IGC population.  The higher value is also consistent with the value
  estimated by Blakeslee \etal (1997), who also used a fainter luminosity.}
This value is very much in line
with the those of non-cD, giant early-type galaxies in the Virgo and
Fornax clusters.

If we assume that all GCs (including IGCs) are part of the NGC~4874
system and that we 
are not missing any luminosity from the galaxy, then this would give the
specific frequency within 520~kpc a higher value of
$S_N=11.4\pm0.2$, a value similar to those measured for some cD
galaxies.  Another 
interpretation, which we discuss later, is that the specific frequency
of the system is the lower, more normal value, but that the IGCs are
tracing a large amount of intracluster star light that is unaccounted for.

\subsection{Comparison to NGC~4874 Surface Brightness Profile}
\label{sec:sbp}

\begin{figure}
\epsscale{1.1}
\plotone{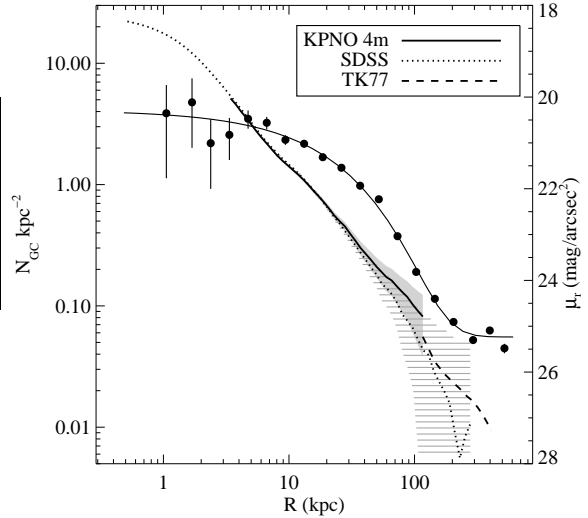}
\caption{The radial distribution of GCs centered on NGC~4874 (black
  dots) compared to the surface brightness profile of field
  star light around NGC~4874.  Three different sources are used for
  the surface brightness profiles: KPNO (solid), SDSS (dotted), Thuan
  \& Kormendy (1977, dashed).  The shaded and striped regions
  represent a change in sky determination of $\pm2\%$ for the KPNO and
  SDSS profiles, respectively.  The GC radial surface number density
  profile and the field star surface brightness profile do not exhibit
  similar shapes at either small or large radii.  The surface
  brightness measurements at large radii, however, are entirely
  dependent on an accurate measure of the sky brightness.
  \label{fig:sbprof}}
\end{figure}

A relevant comparison for the GCs is to the surface brightness profile of the
field star light of NGC~4874.  In Figure~\ref{fig:sbprof}, we plot the
light profile in circular apertures around NGC~4874 from two
independent data sets (with arbitrary normalization).  As mentioned in
the previous section, the first is from
measurements using SDSS 
$r$-band imaging (J.\ Lucey, in preparation), and the second uses imaging
from the Mosaic-I camera on the KPNO 4-meter
telescope (R.\ Marzke, private communication).  Both profiles are in good
agreement in the inner regions ($R<20$~kpc), but start to diverge in the
outer regions due to differences in the sky measurements.  The
difference between the two profiles is at the level of 2\% of the
sky.  

In the regions beyond 100~kpc, we
also plot the surface brightness profile of the ``intracluster
background light'' in the Coma cluster as determined photographically
by Thuan \& Kormendy (1977), transforming from $\mathcal{G}$ to $r$
magnitudes using an offset of $(\mathcal{G}-r)=0.37$~mag, based on
a $(B-V)=0.7$~mag, their published transformation from $\mathcal{G}$
to Thuan-Gunn $r$ (Thuan \& Gunn 1976), and then an offset to SDSS $r$
(Fukugita, Shimasaku, Ichikawa 1995).  This profile appears to match
the SDSS photometry at $R\sim100$~kpc but then continues with a
shallower slope. 

It is clear that at large radii ($R>100$~kpc), the determination of
the sky is crucial to 
the measurement of the ICL.  To illustrate this, we shade in
the regions corresponding to a change in sky determination of $\pm2\%$
of the sky level around the KPNO and SDSS measured profiles.  Although
there is no evidence for a ``break'' in the measured surface brightness
profiles akin to what we see in the GCs, the
surface brightness profile in the regions where
IGCs dominate GC counts is entirely dependent on the determination of
the sky level to better than 1\% and thus is difficult to quantify. 

We can use these profiles to calculate the ``local'' specific
frequency of the outer GCs, although any calculation is highly
uncertain due to sky subtraction for the surface photometry.  
Nevertheless, we can take these surface brightness profiles at face
value to see if the calculated values are reasonable.  Assuming
that the surface brightness at $R=200$~kpc is $\mu_r\approx26.2$~mag
(following the 
Thuan \& Kormendy profile), and using $V-r=0.2$~mag for old metal-poor
stellar populations, then $\mu_V(200\ {\rm kpc})\approx26.4$~mag.  Given the IGC
surface density at these radii (46~arcmin$^{-2}$), we estimate
$S_N(200\ {\rm kpc}) = 5$.  If we assume the profile derived from SDSS data,
however, then $\mu_V(200\ {\rm kpc})\approx27.5$~mag, resulting in 
$S_N(200\ {\rm kpc}) = 13$.  We emphasize that the surface photometry
is extremely uncertain at these radii, and the upper error bar on 
this number is essentially unconstrained.  The local values of $S_N$ at
these kinds of radii have previously been reported to
be quite high (Tamura \etal 2006 in M87 and Rhode \&
Zepf 2001 for NGC~4486), but those measurements are equally uncertain
for similar reasons.  

In the inner regions, there is a notable divergence
between the GCs and the galaxy light.  The galaxy does not show the
prominent core within 10~kpc that the GC system does, only displaying a
flattening in the profile at a smaller radius.

\subsection{Comparison to the M87 GC System}

\begin{figure}
\epsscale{1.1}
\plotone{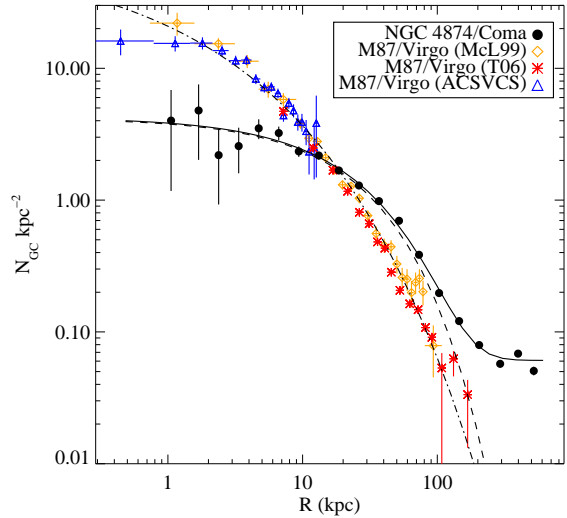}
\caption{The radial distribution of GCs centered on NGC~4874 (black
  dots) compared to the distribution of GCs around the Virgo cD galaxy
  M87.  M87 data is from McLaughlin (1999, McL99, orange diamonds),
  Tamura \etal (2006, T06, red asterisks), and the ACS Virgo Cluster
  Survey (Peng \etal 2008; blue triangles) with a \sersic\ fit to the
  combined data set overplotted (dot-dashed).  The 
  Coma GCs have a much shallower and larger core, as well as an
  inflection where IGCs begin to dominate.  Even with the larger T06
  data set, there is not yet evidence for a profile inflection around
  M87 like what we see in Coma, although the data do not go comparably
  far out in radii.
  \label{fig:m87cmp}}
\end{figure}

Perhaps the most relevant local comparison for the NGC~4874/Coma
Cluster GC system is that of M87 in the Virgo Cluster.  In
Figure~\ref{fig:m87cmp} we show the GC radial surface density profiles
of the two GC systems.  The outer M87 profile is taken from the data of
McLaughlin (1999) and Tamura \etal (2006), while the central regions
of the profile are from the ACS Virgo Cluster Survey (ACSVCS) data shown in
Peng \etal (2008).  We note that the physical resolution of the Coma
{\it HST} data is very competitive with ground-based Virgo
observations ($0\farcs1$ resolution at Coma distance is equivalent to
$0\farcs6$ resolution at Virgo distance), but
obviously cannot match the ACSVCS observations of M87.  None of the
Virgo data sets go as
far out in physical radius as our Coma data, but they still
provide a useful comparison.  The two GC systems profiles are
similar in the range of intermediate radii (20--100~kpc),
but differences appear in the very inner and outer regions.  Most
noticeably, the Coma GC systems displays a very pronounced core within
10~kpc, which does not appear to be present in the Virgo system except
perhaps within 1~kpc.  This
deficit of GCs at the center of NGC~4874 is also evident in the
analysis of Harris \etal (2009).  This core could be the result of
dynamical friction destroying GCs at the center of NGC~4874.  


The core in the GC profile, and the divergence from the M87 GC profile,
is most evident within 10~kpc.  Could this be
due to unaccounted observational incompleteness?  The four innermost
radial bins ($R<3.5$~kpc), have the largest errors and shallowest
observations because 
of the bright galaxy light (completeness of the GCLF is $\approx 25\%$
in these bins).  There are multiple reasons, however, why we believe
these lower surface densities to be real.  The radius at which the core
becomes apparent, $20\arcsec$, is large for $HST$ imaging.  Even
excluding the inner $10\arcsec$ (4.8~kpc), the difference in slope
between the two GC profiles is still apparent.  Also, the completeness
tests we apply also take into account incompleteness due to imperfect
profile subtraction, and thus is a true  measure of the completeness
in these radial annuli.  In order to turn the M87 GC 
profile into the Coma GC profile through a systematic overestimation of
the completeness, the completeness would have to be overestimated by
nearly an order of magnitude.  Lastly, this deficit of GCs relative to
the galaxy light profile was also independently found by Harris \etal
(2009) using HST/WFPC2 data (see their Figure~6). 

In the outer regions the M87 profile follows the single-\sersic\ fit out
to the limits of the data.  The Virgo data, however, only reaches a
radius of 130~kpc, and thus would not be sensitive to the kind of IGC
population we see in Coma.  In fact, if the Coma data had the same
physical radial extent, it would have been very difficult to detect
the IGC population.  More deep, wide-field imaging of the area around
M87, such as the Next Generation Virgo
Survey\footnote{The Next Generation Virgo Survey (NGVS) is a Large
  Program with the Canada-France-Hawaii Telescope.}, will be
necessary to detect or place more stringent limits on a population of
IGCs in Virgo.  

\begin{figure}
\epsscale{1.1}
\plotone{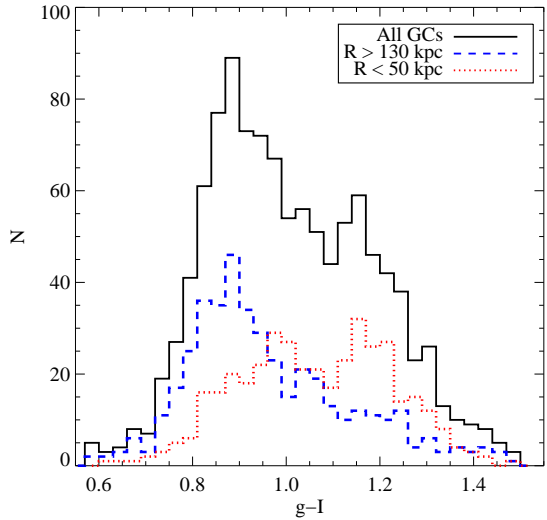}
\caption{The color distributions of all GC candidates in unmasked
  regions with $I<25$ (black
  solid line), a predominantly IGC subsample with $R>130$~kpc (blue
  dashed line), and a predominantly galactic GC subsample
  with $R<50$~kpc (red dotted line).  The total distribution exhibits
  a bimodality typical for extragalactic GC systems, as does the IGC
  sample.  For the IGCs, blue GCs outnumber red GCs by a ratio of
  4:1.  Only in the inner regions, where the GC system of NGC~4874 is
  dominant, does the number of red GCs compare to the number of blue
  GCs.  The color distribution of the inner GCs is also bimodal, but with
  the blue population having a much redder color.
  \label{fig:colordist}}
\end{figure}

\subsection{GC Color Distributions}

For old GCs, the broadband color is an indicator of metallicity.
The color distributions of extragalactic GC systems have been studied
extensively with HST (e.g., Larsen \etal 2001; Peng \etal 2006), and
are often bimodal in nature (Gebhardt \& Kissler-Patig 1999),
especially in massive early-type
galaxies.  In Figure~\ref{fig:colordist}, we plot the color
distribution of bright GCs (applying a magnitude limit of $g<25$~mag for
higher S/N) in the
unmasked regions of the Coma Cluster core.  The color distribution of
all GCs shows the 
typical bimodality seen in extragalactic GC systems, displaying a prominent
peak of blue (metal-poor) GCs with $(g-I)\approx0.9$, and a red
(metal-rich) peak with $(g-I)\approx 1.15$.  

We plot the color
distributions divided by distance from the center of NGC~4874---those
within 50~kpc (galactic GCs) and those outside of 130~kpc
(predominantly IGCs).  We use the Kaye's Mixture Model (KMM; McLachlan \&
Basford 1988; Ashman, Bird, \& Zepf 1994) implementation of the
expectation-maximization (EM) method to fit two Gaussians with
the same standard deviation to the GC color distributions of each sample.  Both
the galactic and intergalactic GCs are much better described by 
bimodal distributions in color than by a single Gaussian with $p$-values
less than $0.001$.  The inner blue GCs, however, have a much 
redder mean color---$(g-I)=0.94$ as opposed to $(g-I)=0.89$ for GCs in
the outer regions---such that
they are nearly merged with the red GCs $(g-I)=1.18$. 
The metal-poor GCs are either quite red or
there is a substantial population of GCs at intermediate color.
It could also be the sign of a significant radial color gradient
within the blue GC subpopulation (e.g., Harris 2009, Liu \etal
2010). This gradient could be in
metallicity or in age, reminiscent of younger metal-poor GCs in Local
Volume dIrr galaxies (Sharina \etal 2005, Georgiev \etal 2008, 2009),
although the colors of the blue GCs are not so blue as to require very
young ages.  The inner regions also have equal numbers of 
blue and red GCs where the fraction of red GCs is $f_{red}=0.51$. 

The generally red colors of the inner GCs was noted
by Harris \etal (2009), who used their WFPC2 data to show that the
inner regions of the GC system had a high ($\gtrsim50\%$) fraction of
red GCs.  Using the larger ACS coverage of our survey, however, we see
that the total GC system around NGC~4874 is dominated by blue GCs.  In
the outer sample plotted in Figure~\ref{fig:colordist}, the IGC
population is dominated by blue GCs with a blue-to-red ratio of 4:1,
($f_{red}=0.2$) although the fraction of red GCs is still significant.
Even measuring only the \sersic\ component of 
the GC subpopulations, 
the NGC~4874 GCs are dominated by blue GCs, with the red GCs being
more spatially concentrated.  The radius at which the surface
density of blue and red GCs are equal is $R\approx40$~kpc.  Thus, like
the giant ellipticals in the Virgo Cluster (Peng \etal 2008), the GC
system of NGC~4874 as well as the IGC population, is dominated by blue GCs.

The mean colors of the blue and red outer GCs are $(g-I)=0.89$ and
1.22, respectively.  The blue peak has a typical color for the GC
systems of galaxies with masses at or below $L^\ast$ (Peng \etal
2006).  The red peak, however, is still fairly red, having nearly the
same mean color as the inner GCs around NGC~4874.  We caution that
some of these red GCs may be from the large metal-rich GC system of
NGC~4889 and other more massive cluster members. 

We can do a simple test to see if the red GCs beyond 130~kpc are
associated with luminous galaxies.  We calculate the distance 
from each GC to the nearest luminous ($M_B<-17$) galaxy, and compare 
the mean shortest distance for the red and blue GCs.  We find that
there is no significant difference between these two populations.  The
median shortest distance for red GCs is $75\arcsec$ and that for blue
GCs is $74\arcsec$, with the biweight Gaussian sigmas of each
distribution being $35\arcsec$.  The spatial behavior of the red and
blue GCs beynd 130~kpc are identical and is consistent with a 
population of IGCs mostly uncorrelated with nearby galaxies.

\section{Discussion}
\subsection{Intracluster Light Inferred From IGCs}
The existence of intracluster starlight and globular clusters has for decades
been considered important, but always difficult to observe.  
Recently, as observations have improved and theory has shown that
intracluster stellar populations are an essential feature of
galaxy-galaxy and galaxy-cluster tidal interactions, there
is increased interest in quantifying 
its properties: total mass, spatial distribution, metallicity, and
kinematics.  We have shown that with {\it HST}, a direct detection of IGCs
is a clean way to measure one component of intracluster stellar
populations in nearby galaxy clusters.

Only a small fraction of the total or stellar mass is in the form of
old globular clusters.  The total mass fraction in GCs appears to be
relatively constant across galaxies and galaxy clusters
(McLaughlin 1999; Blakeslee 1999).  Because of the variation in the
stellar mass-to-light ratio across galaxy mass, however, the specific
frequency (or stellar mass fraction) can vary widely. Massive
ellipticals and dwarf galaxies can have the highest $S_N$, and galaxies
with luminosities around $L^\ast$ have the lowest $S_N$. The
stellar mass fraction in early-type galaxies ranges from $\sim0.2\%$ to
a few percent (Peng \etal 2008).

We do not know the fraction of stellar mass that is in the form of
IGCs, but we can make reasonable assumptions.  One possibility is
that the IGCs originate from low-mass dwarf galaxies that are tidally
disrupted.  The $S_N$ of such a population can vary, and can depend on
the clustercentric radius (Peng \etal 2008), but if we assume that
dwarfs in the cluster center will have high specific
frequencies, then we can reasonably assume $S_{N,IGC}=8$, which is
similar to the $S_N$ values for high-$S_N$ dEs at the center of the Virgo
Cluster (Miller \& Lotz 2007; Seth \etal 2004; Peng \etal 2008),
the GC systems of even lower mass dwarfs (e.g., Puzia \& Sharina
2008), and for Virgo cluster IGCs
(Williams \etal 2007).

Such a value for the specific frequency of the IGC population would
imply a total ICL luminosity within 520~kpc of $M_V=-24.4$~mag, an amount
of star light equal to the whole of NGC~4874.  Spread
out uniformly over this entire area, this luminosity would have a mean
surface brightness of $\mu_{V,ICL}=27 {\rm\ mag\ arcsec}^{-2}$,
which is still challenging for surface photometry, but detectable in
the best observations (Mihos \etal 2005).  If
we assume $M/L_V=1.7$, a value typical of cluster dEs, then this
corresponds to a stellar mass of $\mathcal{M}_{ICL}\approx
9\times10^{11} \ \mathcal{M}_{\odot}$.

Assuming a lower specific frequency like $S_{N,IGC}=1.5$, as is more common for
$L^\ast$ early-type galaxies or low-$S_N$ dEs and dS0s in the
outskirts of the Virgo Cluster, would imply a higher
luminosity for 
the ICL.  Conversely, assuming a very high specific frequency like
that in M87 or the highest-$S_N$ dwarfs, $S_{N,IGC}=12$, would imply a
lower ICL luminosity.  The inferred ICL luminosities, surface
brightnesses, masses, and central ICL fractions for these three assumed values
of $S_{N,IGC}$ are listed in Table~\ref{tab:icl}.  These values
bracket the range of ``local'' $S_N$ we calculate from the outer surface
brightness profiles of NGC~4874 in Section~\ref{sec:sbp}.  Deeper, systematically controlled 
photometry of the intracluster region will be
the best way to set a meaningful limit on the $S_N$ of the IGC population.

\begin{table}
\begin{center}
\caption{ICL Luminosity, Surface Brightness, Mass, and Fraction
  $(R<520 {\rm\ kpc})$ as a Function of $S_{N,IGC}$\label{tab:icl}}
\begin{tabular}{|c|c|c|c|c|}
\hline $S_{N,IGC}$ & $M_{V,ICL}$ & $\mu_{V,ICL}$ & $\mathcal{M}_{ICL}$
& $\frac{\mathcal{L}_{ICL}}{(\mathcal{L}_{N4874+ICL})}$\\
  &  (mag)  &  (mag arcsec$^{-2}$) & ($10^{11} \mathcal{M}_\odot$) & \\
\hline
1.5 & $-26.2$ & 25.2 & $50$ & 0.8 \\
8 & $-24.4$ & 27.0 & $9$ & 0.5 \\
12 & $-24.0$ & 27.4 & $6$ & 0.4 \\
\hline
\end{tabular}
\end{center}
\end{table}

We have also calculated these values using the total luminosity in
GCs, which is more robust than the total number because most of the
luminosity in GCs is brighter than the mean of the GCLF (Harris 1991).
Using typical values for the stellar luminosity fraction (0.2--1.2\%)
from Peng \etal (2008), we arrive at very similar numbers to those in
Table~\ref{tab:icl}.

All of these inferred luminosities and masses are larger
than those previously inferred from direct measurement of the low surface 
brightness ICL in Coma.  Both Gregg \& West (1998) and Adami \etal (2005)
estimate a value of $M_R\approx -22$, which for an old stellar
population is roughly $M_V\approx-21.5$.  However, their values were
the result of combining the luminosities of individual ICL sources.
The 1\% detection limit of the Adami \etal (2005) study is quoted as
$\mu_V=25.8$, which is much shallower than what is required if
$S_{N,IGC}\gtrsim 2$.  It is likely that these previous studies were
sensitive to overdense regions in the ICL, but not to the overall
population of intracluster stars.

\subsection{The Intracluster Fraction}
When considering only the IGC and NGC~4874 GC system (equivalent to
BCG+ICL measurements), we find that the IGCs make up $\sim70\%$ of the
total ``central'' GC system.  Although more difficult, we can also
estimate the IGC fraction for {\it all} GCs in the cluster core,
including the GCs belonging to cluster member galaxies.  We do this in
two ways.  First, we can calculate this number only for those areas
that we have observed.  We assume the IGC surface density to be the fitted
value in Table~\ref{tab:fit}.  We can subtract the surface density
of IGCs from our data and what remains is the ``galactic'' GC
population in the observed areas.  Using this metric, we find that
roughly $\sim30\%$ of the observed GCs belong to the intracluster
population.  This number does not account for the GCs close to
galaxies that were missed by our detection algorithm, so it can be
considered an upper limit over the area observed.

The main drawback of limiting to the observed area, however, is that
we do not observe large portions of the cluster core, including the other
massive elliptical, NGC~4889.  Another method is to apply the GC
system modeling previously described to estimate the total number of
GCs within 520~kpc that belong to galaxies.  When we do this, we find
that the IGCs make up $\sim45\%$ of the GCs in the Coma Cluster core
(this fraction is higher because our observed area includes NGC~4874,
depressing the IGC fraction).
This number is naturally more uncertain because a large fraction of
the galaxies are unobserved.

Given the assumptions in the previous section, we can infer that
approximately half of 
the stellar luminosity associated with the ``central'' system
(NGC~4874 and the ICL) is in the ICL.  Because the ICL is likely to be
more metal-poor and have a lower mass-to-light ratio than the stars at
the center of NGC~4874, this translates to about one-third of the
stellar mass being in the ICL.  We would expect
that the IGC fraction should be higher than the ICL fraction, because
most of the stellar mass is in $L^\ast$ galaxies that have low $S_N$,
whereas the specific frequency of the IGC population is likely to be
high.  This is why we find that the IGC fraction of the ``central''
system is $\sim70\%$, but that the likely ICL fractions in
Table~\ref{tab:icl} are closer to 50\%.

All of these numbers imply a high ICL fraction in the cluster, but are
consistent with expectations from observations and simulations for 
the core of a massive, rich cluster such as Coma.  Seigar \etal (2007)
find an central ICL fraction of 60--80\% around cD galaxies, and
Purcell \etal (2007) expect similar ICL fractions from simulations of
tidal stripping.  

Previous studies of extragalactic GC systems have also suggested that
there is a roughly constant ratio between the total number or mass
of GCs in a system and the total dynamical or baryonic mass (e.g.,
Blakeslee 1999; McLaughlin 1999; Peng \etal 2008; Spitler \etal
2008).  If this is the case, than the IGC fraction may be more
representative of the total mass fraction in the diffuse component of
the galaxy cluster.

\subsection{Scenarios for the Origin of IGCs}

Both theory and simulations suggest that an intracluster stellar
component is a natural result of many physical process that shape
galaxies in the cluster environment.  Galaxy-galaxy tidal interactions
(e.g., Gallagher \& Ostriker 1972; Moore \etal 1996; Stanghellini,
Gonz{\'a}lez-Garc{\'{\i}}a, \& Manchado 2006), tidal forces as
galaxies orbit through the cluster gravitational potential (e.g.,
Merritt 1984; Gnedin 2003), tidal ``preprocessing'' within infalling
galaxy groups (Rudick \etal 2006), tidal destruction of low mass
galaxies (Lopez-Cruz \etal 1997), and tidal tails that escape from
merging galaxies (e.g., Murante \etal 2007)  are all
plausible mechanisms to liberate stars and GCs from the gravitational
potential of their host galaxy. The problem is in trying to
distinguish between these mechanisms, and determine what combination
produces the observed intracluster stellar populations. 

Based on previous ICL studies in Coma, the
specific frequency of the IGC population is not likely to be very low
otherwise its associated ICL would already have been detected at
relatively high S/N.
One straightforward possibility is that the GCs originate mainly from
disrupted dwarf galaxies, which can have high specific frequencies.
Dwarf galaxies, having
low masses and surface mass densities, are prone to be destroyed
by interactions with other galaxies and with the cluster potential.
If we assume $S_{N,IGC}=8$, then the inferred IGC population would
require the disruption of $\sim2300$ dwarf galaxies with $M_V=-16$.


Simulations suggest, however, that the ICL originates from stars in
$L^\ast$ galaxies rather than from dwarfs (Bekki \& Yahagi 2006;
Purcell \etal 2007), and that a significant fraction of the ICL is
first liberated in dynamically cold streams possibly thrown off from
merger remnants (Rudick \etal 2009).  ICL production mechanisms that
depend on galaxy-galaxy interactions appear to be able to produce the
right amount of intracluster mass (early calculations by Gallagher \& Ostriker 
(1972) of stars stripped from $L^\ast$ galaxies during close encounters
predicted that the total ICL mass in the Coma cluster should be
approximately $10^{12} \mathcal{M}_\odot$, a number quite close to what we
infer from the IGCs).  

Galaxy-galaxy harassment, however, is expected to produce a
more centrally concentrated IGC/ICL component than, for instance,
interactions from with the mean cluster tidal field (Merritt 1984),
because the cross section for galaxy-galaxy interaction is strongly
peaked at the cluster center.  We unfortunately cannot determine the
full spatial density profile of the IGCs as our observations only
reach $R=520$~kpc.  What we do measure in the IGC dominated regime
($150<R<520$~kpc), however, is consistent with a flat profile and
shows no sign of being very centrally concentrated.   

Early-type galaxies near $L^\ast$ have a nearly universally low
specific frequency, 
with $S_N\approx 2$ (Peng \etal 2008).  If these kinds of galaxies
were entirely disrupted to supply the necessary IGCs, it would imply a
relatively high ICL surface brightness.  What is more likely is that
it is predominantly the halos of these galaxies that are stripped, and
that the stellar halos have higher local $S_N$.  A prediction of this
scenario would be the presence of a large number of
intermediate-luminosity galaxies in the cluster core that have
relatively few GCs and compact GC systems.  These galaxies might be
detectable as outliers in the color-magnitude relation for early-type galaxies.

The color distribution of the IGCs also provides a clue.  Although the
IGCs are mostly blue, 20\% of the IGCs are still red.  This
appears contrary to the simulations of Bekki \& Yahagi (2006), which
predicted a single metal-poor peak for the GCs with few metal-rich
IGCs.  The red IGCs we 
observe have a mean color consistent with the red GC populations found in the
more massive early-type galaxies rather than in the dwarfs (Larsen
\etal 2001; Peng \etal 2006).  The fraction of red IGCs is also
slightly higher than that in dwarfs.  For Virgo early-type dwarfs with
$M_V\gtrsim -18$, the mean fraction of red GCs is only 10\%, and many have
zero red GCs.  Producing this many red IGCs solely from the disruption
of low mass galaxies would be difficult.  Stripping
of GCs from the halos of more massive galaxies ($M_V\approx -19$)
might be a more natural mechanism for producing the IGC population,
although the stripping could not be too efficient otherwise the ICL
might be too red (c.f., the blue ICL color seen in Virgo by Rudick
\etal 2010).

Mergers of these galaxies could also cause some metal-rich GCs to
escape into intracluster space , although we are not able to
differentiate between merging and stripping.
Dissipational merging, with star formation, could also produce IGCs
within gaseous tidal tails (e.g., Knapp \etal 2006), although for this
mechanism to be important it would have to happen early in the
formation of the cluster.

Although it is likely that dwarf galaxy disruption does play a role in
supplying IGCs, the non-negligible fraction of red IGCs indicates that
stripping and merging of intermediate-mass
galaxies contributes a significant fraction of the population.
Unfortunately, we do not currently have sufficient data to derive the
full spatial profile of the IGC population itself, which would help
determine the originating galaxy population, differentiate between
merging, galaxy-galaxy interactions, and 
galaxy-cluster interactions, and also test whether 
the IGC radial density follows the overall mass density.  More ACS imaging
within the core and at larger radii would be extremely useful for this
purpose.  

\section{Conclusions}

We present the spatial distribution and color distribution of the
globular cluster population in the Coma Cluster core, from imaging
obtained as part of the {\it HST/ACS} Coma Cluster Survey.
\begin{itemize}

\item We discover a large population of GCs that do not appear to be
associated with individual galaxies.  After masking and statistically
subtracting the GC systems of all member galaxies except the central galaxy,
NGC~4874, we find that IGCs dominate the GC surface number density at
galactocentric radii beyond $R>130$~kpc.  These IGCs appear to have a
flat surface density profile out to the extent of our data
($R=520$~kpc).

\item   Using a \sersic\ plus constant model fit to the
radial profile, we estimate that there are $47000\pm1600$
(random) $^{+4000}_{-5000}$ (systematic) IGCs within
520~kpc, and that these make up 70\% of the central Coma Cluster GC
system (NGC~4874+IGCs).
Including the GC systems of cluster members, we still estimate that
IGCs make up 30--45\% of all GCs in the cluster core.  

\item The color distribution of the IGCs is bimodal, with red GC
  fraction of 20\%.  The inner GCs around NGC~4874 also have a
  bimodal color distribution, although with a fairly red metal-poor
  population, and with equal numbers of blue and red GCs. 

\item The reasonably high red fraction (20\%) for the IGCs, and red
  colors of the metal-rich IGCs, suggests that the
  IGC population did not solely originate in dwarf galaxies, but at
  least some part was stripped from the halos of more massive galaxies.  

\item These IGCs trace a large population of stars with an estimated surface
  brightness of $\mu_V\approx 27 {\rm\ mag\ arcsec}^{-2}$ (assuming
  $S_{N,IGC}=8$).  The ICL makes up approximately 
  half of the stellar luminosity and one-third of the stellar mass of
  the central system. The IGCs and ICL are associated with the 
  build up of the central cluster galaxy, and are likely the result of
  the continued growth and evolution of the cluster well after star
  formation has ceased. 

\end{itemize}

\acknowledgments

We thank Naoyuki Tamura for sharing his surface density profile of the
M87 GC system, and Neal Miller for his helpful comments.  We thank the
referee for useful suggestions that improved the manuscript. 

E.~W.~P.\ gratefully acknowledges support from the Peking University
Hundred Talent Fund (985), grant 10873001 from the National
Natural Science Foundation of China, and thanks the Beijing Bookworm
for their hospitality.  T.~H.~P.\ acknowledges support from the
National Research Council of Canada through the Plaskett Research
Fellowship.  D.~C. is supported by STFC rolling grant PP/E001149/1
"Astrophysics Research at LJMU".

Support for program GO-10861 was provided through a grant from the
Space Telescope Science Institute, which is operated by the
Association for Research in Astronomy, Inc., under NASA contract
NAS5-26555.

This research has made use of the NASA/IPAC Extragalactic Database
(NED) which is operated by the Jet Propulsion Laboratory, California
Institute of Technology, under contract with the National Aeronautics
and Space Administration.

This publication makes use of data products from the Sloan Digital
Sky Survey (SDSS).  Funding for SDSS and SDSS-II has
been provided by the Alfred P. Sloan Foundation, the Participating
Institutions, the National Science Foundation, the U.S. Department
of Energy, the National Aeronautics and Space Administration, the
Japanese Monbukagakusho, the Max Planck Society, and the
Higher Education Funding Council for England. The SDSS Web site is
http://www.sdss.org/. 

The SDSS is managed by the Astrophysical Research Consortium (ARC)
for the Participating Institutions. The Participating Institutions
are the American Museum of Natural History, Astrophysical
Institute Potsdam, University of Basel, University of Cambridge,
Case Western Reserve University, The University of Chicago, Drexel
University, Fermilab, the Institute for Advanced Study, the Japan
Participation Group, The Johns Hopkins University, the Joint
Institute for Nuclear Astrophysics, the Kavli Institute for
Particle Astrophysics and Cosmology, the Korean Scientist Group,
the Chinese Academy of Sciences (LAMOST), Los Alamos National
Laboratory, the Max-Planck-Institute for Astronomy (MPIA), the
Max-Planck-Institute for Astrophysics (MPA), New Mexico State
University, Ohio State University, University of Pittsburgh,
University of Portsmouth, Princeton University, the United States
Naval Observatory, and the University of Washington. 

The Digitized Sky Surveys were produced at the Space Telescope Science
Institute under U.S. Government grant NAG W-2166. The images of these
surveys are based on photographic data obtained using the Oschin
Schmidt Telescope on Palomar Mountain and the UK Schmidt
Telescope. The plates were processed into the present compressed
digital form with the permission of these institutions:

The National Geographic Society - Palomar Observatory Sky Atlas
(POSS-I) was made by the California Institute of Technology with
grants from the National Geographic Society. 

The Second Palomar Observatory Sky Survey (POSS-II) was made by the
California Institute of Technology with funds from the National
Science Foundation, the National Geographic Society, the Sloan
Foundation, the Samuel Oschin Foundation, and the Eastman Kodak
Corporation. 

The Oschin Schmidt Telescope is operated by the California Institute
of Technology and Palomar Observatory. 

Facilities: \facility{HST(ACS)}

\appendix
\section{Masking and Modeling of Globular Clusters From Coma Galaxies}
\label{sec:appendix}

\subsection{Generating Masks}
\label{sec:masks}
We first generate masks around bright galaxies in and around all of our fields.
We define masked galaxies to be those with $200\arcsec$ of an ACS field
center, and with $g_0<18$
($M_g<-17$) as observed in the SDSS Data Release 6
(DR6) catalog.  To be conservative, we choose all galaxies that
meet these criteria, although nearly all at these bright magnitudes
are cluster members.  We are required to use a catalog external to our
ACS survey because some large galaxies not observed by ACS, but just
off the edge of a field, can contribute GCs to our observations.

For each galaxy, excluding the two giants NGC~4874 and 4889, we use as
a crude measure of galaxy size the radius that encloses 50\% of the
Petrosian light ($R_{p.50}$).  After experimenting with various mask
sizes, we liberally mask area corresponding to a
circular aperture with a radius of $8 R_{p,50}$.  For a \sersic\
profile with $n=2$, this would mask 99\% of the galaxy light.  Even if
the GC systems were twice the size of the galaxies, this would still
mask $\sim92\%$ of the GC system.  For three of the
larger galaxies, the SDSS sizes are not reliable, so we manually
adjusted the mask size.  For NGC~4889, IC 4045, and 4908, we used
mask radii of 230\arcsec, 68\arcsec, and 160\arcsec, respectively.

\subsection{Simulated GC spatial distributions from galaxies}
\label{sec:simgc}

For NGC~4889 in particular, the GC system is so large that a simple
masking will not eliminate the contribution of its GCs.  This is also
the case for a few other bright galaxies.  There is also the
possibility that smaller but more numerous GC systems from the many
dwarf galaxies in the cluster are contributing to the observed
distribution of GCs.  To account for this, we have simulated the
composite GC distribution we would expect from the known galaxies in Coma.

We start with the Coma Cluster galaxy catalog of Eisenhardt \etal
(2007), who published $UBVRIzJHK_s$ photometry for Coma galaxies
covering a region well-matched to our cluster core mosaic.  For this
purpose, their catalog is superior to the SDSS data because Eisenhardt
\etal (2007) specifically measured photometry of the Coma
galaxies, many of which are too large for the SDSS pipeline to handle
properly.  Their catalog is complete to $M_V<-16$~mag and extends to
$M_V<-14$~mag.  We use their Tables~1 and 2 as the basis for our simulated
GC systems.  We assume that each galaxy in their catalog has a GC
system that is circularly symmetric, and has a radial number density
distribution that follows a \sersic\ (1968) profile with index $n=2$, which
is a good fit to the GC spatial density profiles of the Virgo giant
ellipticals M87 and M49.  The one exception to this is NGC~4889, for
which it seems clear from Figure~\ref{fig:spatialdss} that its
associated GC systems is not spherical.  For this galaxy, we assume
that the spatial density distribution follows a \sersic\ profile along
the geometric mean radius, $\langle r \rangle = \sqrt{ab}$ where $a$ and $b$
are the major and minor axes, respectively.  The ellipticity of
NGC~4889 is $\epsilon = 0.379$ (J{\o}rgenson, Franx, \& Kjaergaard
1992), although we 
find that its GC systems is better described with higher ellipticity,
and use $\epsilon = 0.6$. For this galaxy, we do not need to model its
GCs but can use the observed
radial distribution of GCs from the WFPC2 observations of Harris \etal
(2009).

\begin{figure*}
\plotone{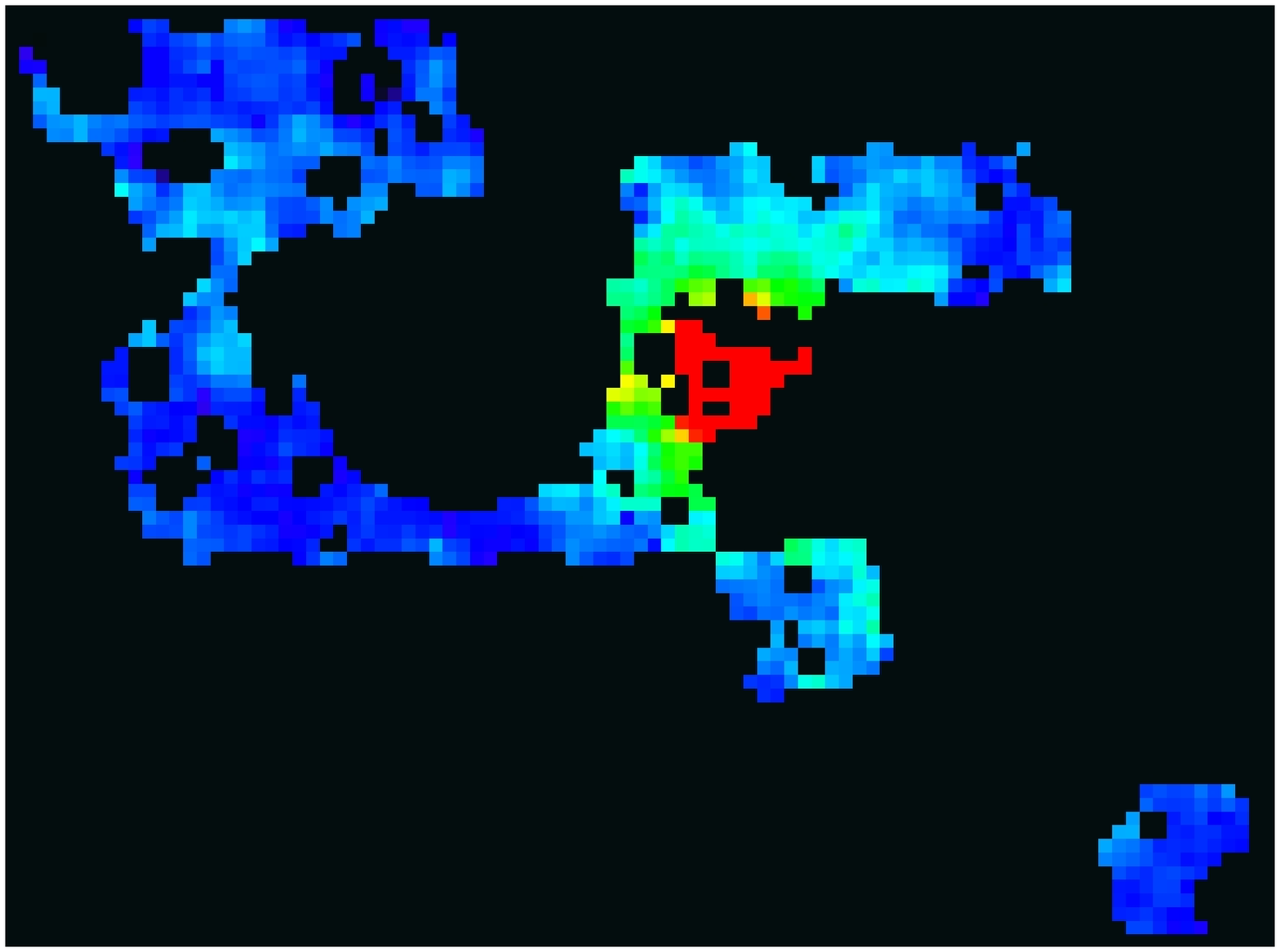}
\caption{
Smoothed spatial distribution of GCs in the Coma Cluster core as in
Figure~\ref{fig:gcgrid} ($30\farcm8\times23\farcm0$,
$900\times670$~kpc), except with 
areas around bright galaxies masked, and modeled contribution from
galactic GC systems subtracted.  The area and surface densities in
this image are used in the radial profile shown in
Figure~\ref{fig:radial}.
\label{fig:sersub}
}
\end{figure*}

We then estimate the GC specific frequency,
$S_N$, for each galaxy based on its absolute $V$ magnitude,
$M_V$.  We use a linear interpolation of the Virgo early-type galaxy
$S_N$ data presented in Table~3 of Peng \etal (2008), giving us an
estimate of the total number of GCs in each galaxy.  Although we do
not make any adjustments for morphological type, most cluster members
in the core are early-types similar to the galaxies analyzed in Peng
\etal (2008), and early-types also generally have higher $S_N$ than
late-types.

Lastly, we determine an effective radius,
$R_e$, for the GC system, also assuming that it scales with galaxy
$M_V$.  We used a 2nd order polynomial fit between $R_e$ and $M_V$ in
which GC systems of more luminous galaxies are larger.  We derived
this relation using the GC systems of 100 Virgo cluster galaxies
observed in the ACS Virgo Cluster Survey (\cote \etal 2004; \jordan
\etal 2009).  This survey observed early-type galaxies with a wide range of
luminosities ($-22<M_B<-15$). The spatial density profiles were fit
with \sersic\ profiles to derive the effective radius of the GC
systems as a function of galaxy luminosity.  The details of
this relation and its derivation will appear in a subsequent paper
(Peng et al, in prep). The spatial extent of GC systems is generally
larger than that of the stars, and is consistent with previous
measurements showing that the local $S_N$ for giant ellipticals
increases with galactocentric radius.  With this estimate of $R_e$, we
can then infer the contribution of GCs from galactic systems as a
function of position in the central mosaic of observations.

Using the same spatial grid shown in Figure~\ref{fig:gcgrid}, we 
estimate the total number of GCs expected in each cell from {\it all} of 
the galaxies in the Eisenhardt \etal (2007) catalog (not
including NGC~4874).  We correct this number for the
total observed area---accounting for unobserved and masked area---and
the local GCLF completeness.  For each cell,
we thus have an estimate of the total number of observed GCs that
could be contributed by the GC systems of known cluster galaxies.

We then use this simulated distribution of GCs to statistically
subtract the contribution from ``galactic'' GCs.  In each cell, we
sample a Poisson
distribution about the expected value, and then assign each GC a
random position in the cell.  Additionally, we mask regions around all
the known galaxies, as described above, with the results shown in
Figure~\ref{fig:sersub}.  We do expect that some galactic GCs will
contribute to
the apparent intergalactic GC population, especially GCs from the
other gE, NGC~4889, but the expected number and spatial distribution of the
galactic GCs (i.e., concentrated around the galaxies) do not
agree with the spatial uniformity and overall large numbers of
GCs observed throughout the cluster core.

We also note that although our modeling of GC systems only extends to
galaxies with $M_V<-14$~mag, this is unlikely to bias our final result.
While it is true that $S_N$ does tend
to rise at lower luminosities, and some individual dEs have high
$S_N$, the mean value is not very high.  As shown by
Lotz \& Miller (2007) and Peng \etal (2008), dEs 
and dSphs have a very wide range of $S_N$, with some being high but many
others quite ordinary or near zero.  The mean $S_N$ for dwarfs in the
magnitude range that concerns us is $\sim2$.  Studies of Local Group
dwarfs show that fainter than a certain luminosity ($M_B\approx-10$),
dSphs do not appear to have GCs.  

\subsection{Systematic Errors From Modeling}
\label{sec:systematics}

The technique we use to account for the unmasked galactic GCs
introduces a systematic uncertainty into our analysis.  We can
quantify this effect by varying the assumed parameters over a wide
range of acceptable values.  We detail this in this section, but our
main point is that {\it our conclusions are not particularly sensitive
  to the specific parameters assumed for the modeling of galactic GC
  systems}, because if Coma early-type galaxies have the 
same range of specific frequencies as Virgo early-type galaxies,
then there are simply not enough galactic GCs to account for all the GCs
that we observe. 


We have produced a
range of plausible galactic GC populations, varying the \sersic\ $n$
index of the assumed profiles, the effective radius, $R_e$, and the
specific frequencies of the galaxies, $S_N$.
By varying the Sersic index from $n=1.0$ to
$n=4.0$ (exponential to de~Vaucouleurs), the inferred number of IGCs
varies from $47000$ to $44000$.  The exponential profile is steepest in
the outer regions and so gives the highest estimate for the IGC
population.  If we assume that all galactic GC systems have the
shallower outer profiles of a de~Vaucouleurs profile, then our
inferred IGC population is smaller by $\sim3000$ GCs.  This effect is only
at the peak-to-peak level of 6--7\%, and is small enough to justify our
choice of a single Sersic index for our GCS models.  A single $n=2$
model gives us an inferred population of $46500$~IGCs, so we take the
systematic error here to be ($^{+500}_{-3,000}$).

By varying the specific frequency, we affect the total number of galactic
GCs.  Our modeling uses the $S_N-L$ relationship from Peng et
al. (2008).  Although there is no evidence that $S_N$ for normal
galaxies is different in the Coma cluster, we can make the
assumption that $S_N$ is systematically different from Virgo by $\pm20\%$.  Doing
so would change the number of IGCs by $\pm2,000$.  This does not
appreciably alter the significance of the detection.  We include this
in the systematic error budget.

We can also test the dependence of inferred IGC numbers on the
effective radii of the GC 
systems.  We can assume that the $R_e$ of GC systems in Coma are a
factor of two larger or smaller for a given galaxy luminosity than
for their counterparts in Virgo (a fairly extreme assumption).  If the
GC systems are larger, the number of unmasked galactic GCs is
increased, and the number of IGCs will go 
down.  Using this assumption in our modeling, the total
number of IGCs changes by $\pm3,000$.  Once again, the overall
effect on the significance of our detection is rather small.

Given that our
systematic errors are comparable to or larger than our Poisson errors
(which are at the level of $\pm1600$~GCs), it is important to include
these in the error budget.  Nevertheless, it is clear that the
details of the modeling does not affect the conclusion that there is a
significant detection of IGCs 
in the Coma core.  By combining the estimated errors
independently and in quadrature,
we estimate that there is a combined systematic error of
$^{+4000}_{-5000}$~IGCs.  The inferred number of IGCs is thus 
$47000\pm1600$~(random)~$^{+4000}_{-5000}$ (systematic).

\subsection{A Test of GC System Modeling}
\label{sec:model test}

There are three outer ACS fields that we do not use for our background
estimate because they are near larger Coma galaxies.  Assuming that
these fields have few or no IGCs, they present the opportunity to test
the validity of our methodology.  This test is not ideal, because 
NGC~4839 is a giant elliptical galaxy that is the 
dominant member of its subgroup, and our assumptions are most valid
when we can average over many galaxies rather than be dominated by
a luminous single galaxy.  Nevertheless, we can apply the
technique used to model the core region and check for consistency.
Similar to  NGC~4889, for which there are published measurements and
no need to infer from scaling relations, we
use the known $S_N$ for NGC~4839 (Mar\'{i}n-Franch \etal 2002),
although for every other parameter and for all other galaxies in the
region, we use our scaling relations.  We take the same approach to
modeling and masking as was done in the core, although since the
Eisenhardt \etal (2007) catalog does not cover this region, we use an equivalent
cut with SDSS.  For each field, after subtracting the modeled
GC population and the estimated background from the total counts, we
should expect to have numbers statistically equivalent to zero.  For
these three fields, we obtain residual counts of $13\pm9$, $3\pm11$, and
$10\pm13$, where the errors do not include the systematic errors as
discussed above.  We can assume that the systematic errors contribute
roughly equally (or more) to the error budget.  These numbers are
satisfyingly consistent with zero, within the random and systematic
errors.  The predicted values do skew positive (with large errors), but
this is likely a reflection of our 
systematic error, and is consistent with what we discuss in
Appendix~\ref{sec:systematics}.  Over three fields, the residual
is marginally significant over random errors, $26\pm19$ or when
corrected for the full GCLF, $0.007\pm0.005$~kpc$^{-2}$.  This is
still $8\times$ smaller than our measured IGC surface density, so any
systematic residual at this level would not affect the significance of
our measurement.  Even if we were to adjust our model to
increase the predicted number of GCs so as to match these fields, the
impact on our 
result and the IGCs numbers would be small and within our quoted
errors. Thus, we are reassured that our GC subtraction procedure and
estimation of our uncertainties is robust.


{}

\clearpage


\begin{thebibliography}{}

\bibitem[Adami et 
al.(2005)]{2005A&A...429...39A} Adami, C., et al.\ 2005, \aap, 429, 39 

%
\bibitem[Arnaboldi et al.(2004)]{2004ApJ...614L..33A} Arnaboldi, M., 
Gerhard, O., Aguerri, J.~A.~L., Freeman, K.~C., Napolitano, N.~R., Okamura, 
S., \& Yasuda, N.\ 2004, \apjl, 614, L33 

\bibitem[Arnaboldi et al.(2007)]{2007PASJ...59..419A} Arnaboldi, M., 
Gerhard, O., Okamura, S., Kashikawa, N., Yasuda, N., 
\& Freeman, K.~C.\ 2007, \pasj, 59, 419
\bibitem[Ashman et al.(1994)]{1994AJ....108.2348A} Ashman, K.~M., Bird, 
C.~M., \& Zepf, S.~E.\ 1994, \aj, 108, 2348

%
%
%
%

\bibitem[Balcells et al.(2010)]{2010ApJ...submitted} Balcells, M., et
  al.\ 2010, \apj, submitted

\bibitem[Bassino et al.(2003)]{2003A&A...399..489B} Bassino, L.~P., Cellone, S.~A., Forte, J.~C., \& Dirsch, B.\ 2003, \aap, 399, 489 

\bibitem[Bassino et al.(2006)]{2006A&A...451..789B} Bassino, L.~P., Faifer, F.~R., Forte, J.~C., Dirsch, B., Richtler, T., Geisler, D., \& Schuberth, Y.\ 2006, \aap, 451, 789 

%
\bibitem[Beasley et al.(2008)]{2008MNRAS.386.1443B} Beasley, M.~A., 
Bridges, T., Peng, E., Harris, W.~E., Harris, G.~L.~H., Forbes, D.~A., 
\& Mackie, G.\ 2008, \mnras, 386, 1443 

\bibitem[Bergond et 
al.(2007)]{2007A&A...464L..21B} Bergond, G., et al.\ 2007, \aap, 464, L21 

%
%

\bibitem[Bekki 
\& Yahagi(2006)]{2006MNRAS.372.1019B} Bekki, K., \& Yahagi, H.\ 2006, \mnras, 372, 1019 

%
%
%
%

\bibitem[Blakeslee(1995)]{1995ApJ...442..579B} Blakeslee, J.~P. \&
  Tonry, J.~T.\ 1995, 442, 579

\bibitem[Blakeslee(1997)]{1997ApJ...481L..59B} Blakeslee, J.~P.\ 1997, 
\apjl, 481, L59

\bibitem[Blakeslee et al.(1997)]{1997AJ....114..482B} Blakeslee, J.~P., 
Tonry, J.~L., \& Metzger, M.~R.\ 1997, \aj, 114, 482

\bibitem[Blakeslee(1999)]{1999AJ....118.1506B} Blakeslee, J.~P.\ 1999, \aj, 
118, 1506

%
%
%
%
%
%
%

\bibitem[Calc{\'a}neo-Rold{\'a}n et al.(2000)]{2000MNRAS.314..324C} 
Calc{\'a}neo-Rold{\'a}n, C., Moore, B., Bland-Hawthorn, J., Malin, D., 
\& Sadler, E.~M.\ 2000, \mnras, 314, 324 

%

\bibitem[Carter et al.(2008)]{2008ApJS..176..424C} Carter, D., et al.\ 
2008, \apjs, 176, 424 

\bibitem[Castro-Rodrigu{\'e}z et 
al.(2009)]{2009A&A...507..621C} Castro-Rodrigu{\'e}z, N., Arnaboldi, M., Aguerri, J.~A.~L., Gerhard, O., Okamura, S., Yasuda, N., \& Freeman, K.~C.\ 2009, \aap, 507, 621 

%
%
%
%
%

\bibitem[Chiboucas et al.(2010)]{2010arXiv1009.3950C} Chiboucas, K., et 
al.\ 2010, "A Universe of Dwarf Galaxies", Lyon, France, June 14-18, 2010,
(arXiv:1009.3950)

\bibitem[Collins et al.(2009)]{2009Natur.458..603C} Collins, C.~A., et al.\ 
2009, \nat, 458, 603 

%
%
%

\bibitem[C{\^ o}t{\' e} et al.(2004)]{2004ApJS..153..223C} C{\^ o}t{\' e}, 
P., et al.\ 2004, \apjs, 153, 223

%
%
%

\bibitem[Da Rocha 
\& Mendes de Oliveira(2005)]{2005MNRAS.364.1069D} Da Rocha, C., \& Mendes de Oliveira, C.\ 2005, \mnras, 364, 1069 

%
\bibitem[De Lucia \& Blaizot(2007)]{2007MNRAS.375....2D} De Lucia, G., \& 
Blaizot, J.\ 2007, \mnras, 375, 2 

\bibitem[de Vaucouleurs 
\& de Vaucouleurs(1970)]{1970ApL.....5..219D} de Vaucouleurs, G., \& de Vaucouleurs, A.\ 1970, \aplett, 5, 219 

%
%

\bibitem[Demarco et
al.(2003)]{2003A&A...407..437D} Demarco, R., Magnard, F., Durret, F., \& M{\'a}\
rquez, I.\ 2003, \aap, 407, 437

%
%
%
%

\bibitem[Doherty et 
al.(2009)]{2009A&A...502..771D} Doherty, M., et al.\ 2009, \aap, 502, 771 

\bibitem[Drinkwater et al.(2003)]{2003Natur.423..519D} Drinkwater, M.~J., 
Gregg, M.~D., Hilker, M., Bekki, K., Couch, W.~J., Ferguson, H.~C., Jones, 
J.~B., \& Phillipps, S.\ 2003, \nat, 423, 519 

%
%
\bibitem[Durrell et al.(2002)]{2002ApJ...570..119D} Durrell, P.~R., 
Ciardullo, R., Feldmeier, J.~J., Jacoby, G.~H., 
\& Sigurdsson, S.\ 2002, \apj, 570, 119 

\bibitem[Eisenhardt et al.(2007)]{2007ApJS..169..225E} Eisenhardt, P.~R., 
De Propris, R., Gonzalez, A.~H., Stanford, S.~A., Wang, M., 
\& Dickinson, M.\ 2007, \apjs, 169, 225 

%
%
%
%
\bibitem[Feldmeier et al.(1998)]{1998ApJ...503..109F} Feldmeier, J.~J., 
Ciardullo, R., \& Jacoby, G.~H.\ 1998, \apj, 503, 109 

\bibitem[Feldmeier et al.(2002)]{2002ApJ...575..779F} Feldmeier, J.~J., 
Mihos, J.~C., Morrison, H.~L., Rodney, S.~A., 
\& Harding, P.\ 2002, \apj, 575, 779

\bibitem[Feldmeier et al.(2004a)]{2004ApJ...609..617F} Feldmeier, J.~J., 
Mihos, J.~C., Morrison, H.~L., Harding, P., Kaib, N., 
\& Dubinski, J.\ 2004, \apj, 609, 617 

\bibitem[Feldmeier et al.(2004b)]{2004ApJ...615..196F} Feldmeier, J.~J., 
Ciardullo, R., Jacoby, G.~H., \& Durrell, P.~R.\ 2004, \apj, 615, 196 

\bibitem[Ferguson et al.(1998)]{1998Natur.391..461F} Ferguson, H.~C., 
Tanvir, N.~R., \& von Hippel, T.\ 1998, \nat, 391, 461

%
%
%
%
%
%

\bibitem[Fruchter 
\& Hook(2002)]{2002PASP..114..144F} Fruchter, A.~S., \& Hook, R.~N.\ 2002, \pasp, 114, 144 

\bibitem[Fukugita et al.(1995)]{1995PASP..107..945F} Fukugita, M., 
Shimasaku, K., \& Ichikawa, T.\ 1995, \pasp, 107, 945 

\bibitem[Gal-Yam et al.(2003)]{2003AJ....125.1087G} Gal-Yam, A., Maoz, D., 
Guhathakurta, P., \& Filippenko, A.~V.\ 2003, \aj, 125, 1087 

%

\bibitem[Gebhardt \& Kissler-Patig(1999)]{1999AJ....118.1526G} Gebhardt, 
K., \& Kissler-Patig, M.\ 1999, \aj, 118, 1526

%

\bibitem[Georgiev et al.(2008)]{2008AJ....135.1858G} Georgiev, I.~Y., 
Goudfrooij, P., Puzia, T.~H., \& Hilker, M.\ 2008, \aj, 135, 1858 

\bibitem[Georgiev et al.(2009)]{2009MNRAS.392..879G} Georgiev, I.~Y., 
Puzia, T.~H., Hilker, M., \& Goudfrooij, P.\ 2009, \mnras, 392, 879 

\bibitem[Gerhard et 
al.(2007)]{2007A&A...468..815G} Gerhard, O., Arnaboldi, M., Freeman, K.~C., Okamura, S., Kashikawa, N., \& Yasuda, N.\ 2007, \aap, 468, 815 

\bibitem[Gnedin(2003)]{2003ApJ...582..141G} Gnedin, O.~Y.\ 2003, \apj, 582, 
141 

\bibitem[Gonzalez et al.(2005)]{2005ApJ...618..195G} Gonzalez, A.~H., 
Zabludoff, A.~I., \& Zaritsky, D.\ 2005, \apj, 618, 195 

\bibitem[Gonzalez et al.(2007)]{2007ApJ...666..147G} Gonzalez, A.~H., 
Zaritsky, D., \& Zabludoff, A.~I.\ 2007, \apj, 666, 147 

%
%


\bibitem[Gregg 
\& West(1998)]{1998Natur.396..549G} Gregg, M.~D., \& West, M.~J.\ 1998, \nat, 396, 549 

\bibitem[Gregg et al.(2009)]{2009AJ....137..498G} Gregg, M.~D., et al.\ 
2009, \aj, 137, 498 

\bibitem[Ha{\c s}egan et al.(2005)]{2005ApJ...627..203H} Ha{\c s}egan, M., 
et al.\ 2005, \apj, 627, 203

\bibitem[Hammer et al.(2010)]{2010arXiv1005.3300H} Hammer, D., et al.\ 
2010, \apj, submitted, arXiv:1005.3300


\bibitem[Harris \& van den Bergh(1981)]{1981AJ.....86.1627H} Harris, W.~E., 
\& van den Bergh, S.\ 1981, \aj, 86, 1627

\bibitem[Harris(1991)]{1991ARA&A..29..543H} Harris, W.~E.\ 1991, \araa, 29, 
543

%
%
\bibitem[Harris et al.(2000)]{2000ApJ...533..137H} Harris, W.~E., 
Kavelaars, J.~J., Hanes, D.~A., Hesser, J.~E., 
\& Pritchet, C.~J.\ 2000, \apj, 533, 137 

%

\bibitem[Harris et al.(2009)]{2009AJ....137.3314H} Harris, W.~E., 
Kavelaars, J.~J., Hanes, D.~A., Pritchet, C.~J., 
\& Baum, W.~A.\ 2009, \aj, 137, 3314 

\bibitem[Harris(2009)]{2009ApJ....703.939} Harris, W.~E.\ 2009, \apj, 703, 939 

\bibitem[Hilker et 
al.(2007)]{2007A&A...463..119H} Hilker, M., Baumgardt, H., Infante, L., Drinkwater, M., Evstigneeva, E., \& Gregg, M.\ 2007, \aap, 463, 119 

%
%

\bibitem[Jedrzejewski(1987)]{1987MNRAS.226..747J} Jedrzejewski, R.~I.\ 
1987, \mnras, 226, 747 

\bibitem[Jord{\'a}n et al.(2003)]{2003AJ....125.1642J} Jord{\'a}n, A., 
West, M.~J., C{\^o}t{\'e}, P., \& Marzke, R.~O.\ 2003, \aj, 125, 1642 


\bibitem[Jord{\' a}n et al.(2004)]{2004ApJS..154..509J} Jord{\' a}n, A., et 
al.\ 2004, \apjs, 154, 509

%
%
\bibitem[Jord{\'a}n et al.(2006)]{2006ApJ...651L..25J} Jord{\'a}n, A., et 
al.\ 2006, \apjl, 651, L25

\bibitem[Jord{\'a}n et al.(2007)]{2007ApJS..171..101J} Jord{\'a}n, A., et 
al.\ 2007, \apjs, 171, 101


\bibitem[Jord{\'a}n et al.(2009)]{2009ApJS..180...54J} Jord{\'a}n, A., et 
al.\ 2009, \apjs, 180, 54 

\bibitem[Jorgensen et 
al.(1992)]{1992A&AS...95..489J} Jorgensen, I., Franx, M., \& Kjaergaard, P.\ 1992, \aaps, 95, 489 


%
\bibitem[Kavelaars et al.(2000)]{2000ApJ...533..125K} Kavelaars, J.~J., 
Harris, W.~E., Hanes, D.~A., Hesser, J.~E., 
\& Pritchet, C.~J.\ 2000, \apj, 533, 125 

%
%

\bibitem[Knapp et al.(2006)]{2006AJ....131..859K} Knapp, G.~R., et al.\ 
2006, \aj, 131, 859

%
\bibitem[Koekemoer et al.(2002)]{2002hstc.conf..339K} Koekemoer, A.~M., 
Fruchter, A.~S., Hook, R.~N., \& Hack, W.\ 2002, The 2002 HST Calibration 
Workshop : Hubble after the Installation of the ACS and the NICMOS Cooling 
System, Proceedings of a Workshop held at the Space Telescope Science 
Institute, Baltimore, Maryland, October 17 and 18, 2002.~ Edited by 
Santiago Arribas, Anton Koekemoer, and Brad Whitmore.~Baltimore, MD: Space 
Telescope Science Institute, 2002., p.339, 339

\bibitem[Kormendy 
\& Bahcall(1974)]{1974AJ.....79..671K} Kormendy, J., \& Bahcall, J.~N.\ 1974, \aj, 79, 671 

%
%
\bibitem[Krick 
\& Bernstein(2007)]{2007AJ....134..466K} Krick, J.~E., \& Bernstein, R.~A.\ 2007, \aj, 134, 466 


%
%
%
%
\bibitem[Larsen et al.(2001)]{2001AJ....121.2974L} Larsen, S.~S., Brodie, 
J.~P., Huchra, J.~P., Forbes, D.~A., \& Grillmair, C.~J.\ 2001, \aj, 121, 
2974

\bibitem[Lee et al.(2010)]{2010Sci...328..334L} Lee, M.~G., Park, H.~S., 
\& Hwang, H.~S.\ 2010, Science, 328, 334 

\bibitem[Lin 
\& Mohr(2004)]{2004ApJ...617..879L} Lin, Y.-T., \& Mohr, J.~J.\ 2004,
\apj, 617, 879 

\bibitem[Liu et al.(2010)]{2010arXiv1012.2634L} Liu, C., Peng, E.~W., 
Jord\'{a}n, A., Ferrarese, L., Blakeslee, J.~P., C\^{o}t\'{e}, P., 
\& Mei, S.\ 2010, \apj, accepted (arXiv:1012.2634)

\bibitem[Lopez-Cruz et al.(1997)]{1997ApJ...475L..97L} Lopez-Cruz, O., Yee, 
H.~K.~C., Brown, J.~P., Jones, C., \& Forman, W.\ 1997, \apjl, 475, L97 

%
%

\bibitem[Madrid et al.(2010)]{2010arXiv1009.3023M} Madrid, J.~P., et al.\ 
2010, \apj, 722, 1707

\bibitem[Mattila(1977)]{1977A&A....60..425M} Mattila, K.\ 1977, \aap, 60, 425 

\bibitem[Mar{\'{\i}}n-Franch 
\& Aparicio(2002)]{2002ApJ...568..174M} Mar{\'{\i}}n-Franch, A., \& Aparicio, A.\ 2002, \apj, 568, 174 

\bibitem[Mar{\'{\i}}n-Franch 
\& Aparicio(2003)]{2003ApJ...585..714M} Mar{\'{\i}}n-Franch, A., \& Aparicio, A.\ 2003, \apj, 585, 714 

\bibitem[Matthews et al.(1964)]{1964ApJ...140...35M} Matthews, T.~A., 
Morgan, W.~W., \& Schmidt, M.\ 1964, \apj, 140, 35 

%

\bibitem[McLachlan \& Basford(1988)]{1988MB} McLachlan, G. J., \&
  Basford, K.\ E.\ 1988, Mixture Models: Inference and Application to
  Clustering (New York: M.\ Dekker)

\bibitem[McLaughlin(1999)]{1999AJ....117.2398M} McLaughlin, D.~E.\ 1999,
\aj, 117, 2398

%
%
%
%

\bibitem[Melnick et al.(1977)]{1977MNRAS.180..207M} Melnick, J., Hoessel, 
J., \& White, S.~D.~M.\ 1977, \mnras, 180, 207 

\bibitem[Mendez et al.(1997)]{1997ApJ...491L..23M} Mendez, R.~H., Guerrero, 
M.~A., Freeman, K.~C., Arnaboldi, M., Kudritzki, R.~P., Hopp, U., 
Capaccioli, M., \& Ford, H.\ 1997, \apjl, 491, L23 


\bibitem[Merritt(1984)]{1984ApJ...276...26M} Merritt, D.\ 1984, \apj, 276, 
26

\bibitem[Merritt et al.(2006)]{2006AJ....132.2685M} Merritt, D., Graham, 
A.~W., Moore, B., Diemand, J., \& Terzi{\'c}, B.\ 2006, \aj, 132, 2685 


\bibitem[Mieske et 
al.(2008)]{2008A&A...487..921M} Mieske, S., et al.\ 2008, \aap, 487, 921 

\bibitem[Mihos et al.(2005)]{2005ApJ...631L..41M} Mihos, J.~C., Harding, 
P., Feldmeier, J., \& Morrison, H.\ 2005, \apjl, 631, L41 

\bibitem[Mihos et al.(2009)]{2009ApJ...698.1879M} Mihos, J.~C., Janowiecki, 
S., Feldmeier, J.~J., Harding, P., \& Morrison, H.\ 2009, \apj, 698, 1879 

%
\bibitem[Miller \& Lotz(2007)]{2007ApJ...670.1074M} Miller, B.~W., \& Lotz, J.~M.\ 2007, \apj, 670, 1074 
%

\bibitem[Monaco et al.(2006)]{2006ApJ...652L..89M} Monaco, P., Murante, G., 
Borgani, S., \& Fontanot, F.\ 2006, \apjl, 652, L89

%
%
\bibitem[Moore et al.(2006)]{2006MNRAS.368..563M} Moore, B., Diemand, J., 
Madau, P., Zemp, M., \& Stadel, J.\ 2006, \mnras, 368, 563

\bibitem[Murante et al.(2007)]{2007MNRAS.377....2M} Murante, G., Giovalli, 
M., Gerhard, O., Arnaboldi, M., Borgani, S., 
\& Dolag, K.\ 2007, \mnras, 377, 2 


\bibitem[Neill et al.(2005)]{2005ApJ...618..692N} Neill, J.~D., Shara, 
M.~M., \& Oegerle, W.~R.\ 2005, \apj, 618, 692 

\bibitem[Oemler(1976)]{1976ApJ...209..693O} Oemler, A., Jr.\ 1976, \apj, 
209, 693 

\bibitem[Okamura et al.(2002)]{2002PASJ...54..883O} Okamura, S., et al.\ 
2002, \pasj, 54, 883 

%
\bibitem[Peng et al.(2004)]{2004ApJ...602..705P} Peng, E.~W., Ford, H.~C., 
\& Freeman, K.~C.\ 2004, \apj, 602, 705 

\bibitem[Peng et al.(2006)]{2006ApJ...639...95P} Peng, E.~W., et al.\ 2006, 
\apj, 639, 95


\bibitem[Peng et al.(2008)]{2008ApJ...681...197P} Peng, E.~W., et al.\ 2008, 
\apj, 681, 197

\bibitem[Peng et al.(2009)]{2009ApJ...703...42P} Peng, E.~W., et al.\ 2009, 
\apj, 703, 42

\bibitem[Price et al.(2009)]{2009MNRAS.397.1816P} Price, J., et al.\ 2009, 
\mnras, 397, 1816 

\bibitem[Puchwein et al.(2010)]{2010MNRAS.tmp..788P} Puchwein, E., 
Springel, V., Sijacki, D., \& Dolag, K.\ 2010, \mnras, 788 

\bibitem[Purcell et al.(2007)]{2007ApJ...666...20P} Purcell, C.~W., 
Bullock, J.~S., \& Zentner, A.~R.\ 2007, \apj, 666, 20 


%

\bibitem[Puzia et 
al.(2005)]{2005A&A...439..997P} Puzia, T.~H., Kissler-Patig, M., Thomas, D., Maraston, C., Saglia, R.~P., Bender, R., Goudfrooij, P., \& Hempel, M.\ 2005, \aap, 439, 997 

\bibitem[Puzia 
\& Sharina(2008)]{2008ApJ...674..909P} Puzia, T.~H., \& Sharina, M.~E.\ 2008, \apj, 674, 909 

\bibitem[Rhode \& Zepf(2001)]{2001AJ....121..210R} Rhode, K.~L., \& Zepf, 
S.~E.\ 2001, \aj, 121, 210

%
%
\bibitem[Rhode et al.(2007)]{2007AJ....134.1403R} Rhode, K.~L., Zepf, 
S.~E., Kundu, A., \& Larner, A.~N.\ 2007, \aj, 134, 1403 

%
%

\bibitem[Rudick et al.(2006)]{2006ApJ...648..936R} Rudick, C.~S., Mihos, 
J.~C., \& McBride, C.\ 2006, \apj, 648, 936 

\bibitem[Rudick et al.(2009)]{2009ApJ...699.1518R} Rudick, C.~S., Mihos, 
J.~C., Frey, L.~H., \& McBride, C.~K.\ 2009, \apj, 699, 1518 

\bibitem[Rudick et al.(2010)]{2010ApJ...720..569R} Rudick, C.~S., Mihos, 
J.~C., Harding, P., Feldmeier, J.~J., Janowiecki, S., 
\& Morrison, H.~L.\ 2010, \apj, 720, 569 


%
%

\bibitem[Schlegel et al.(1998)]{1998ApJ...500..525S} Schlegel, D.~J., 
Finkbeiner, D.~P., \& Davis, M.\ 1998, \apj, 500, 525


\bibitem[Schuberth et 
al.(2008)]{2008A&A...477L...9S} Schuberth, Y., Richtler, T., Bassino, L., \& Hilker, M.\ 2008, \aap, 477, L9 

\bibitem[Scoville et al.(2007)]{2007ApJS..172...38S} Scoville, N., et al.\ 
2007, \apjs, 172, 38


\bibitem[Seigar et al.(2007)]{2007MNRAS.378.1575S} Seigar, M.~S., Graham, 
A.~W., \& Jerjen, H.\ 2007, \mnras, 378, 1575 

\bibitem[Sersic(1968)]{1968adga.book.....S} Sersic, J.~L.\ 1968, Cordoba, 
Argentina: Observatorio Astronomico, 1968

\bibitem[Seth et al.(2004)]{2004AJ....127..798S} Seth, A., Olsen, K., 
Miller, B., Lotz, J., \& Telford, R.\ 2004, \aj, 127, 798

\bibitem[Sharina et 
al.(2005)]{2005A&A...442...85S} Sharina, M.~E., Puzia, T.~H., \& Makarov, D.~I.\ 2005, \aap, 442, 85 

\bibitem[Sirianni et al.(2005)]{2005PASP..117.1049S} Sirianni, M.,
  Jee, M.J., Benítez, N., Blakeslee, J.P., Martel, A.R., Meurer, G.,
  Clampin, M., De Marchi, G., Ford, H.C., Gilliland, R., Hartig, G.F.,
  Illingworth, G.D., Mack, J., \& McCann, W.J. 2005, \pasp, 117, 1049

%
%
%
\bibitem[Spitler 
\& Forbes(2009)]{2009MNRAS.392L...1S} Spitler, L.~R., \& Forbes, D.~A.\ 2009, \mnras, 392, L1 

%
%

\bibitem[Stanghellini et al.(2006)]{2006ApJ...644..843S} Stanghellini, L., 
Gonz{\'a}lez-Garc{\'{\i}}a, A.~C., \& Manchado, A.\ 2006, \apj, 644, 843 

\bibitem[Stetson(1987)]{1987PASP...99..191S} Stetson, P.~B.\ 1987, \pasp, 
99, 191 



\bibitem[Tamura et al.(2006)]{2006MNRAS.373..601T} Tamura, N., Sharples, 
R.~M., Arimoto, N., Onodera, M., Ohta, K., 
\& Yamada, Y.\ 2006, \mnras, 373, 601 

\bibitem[Theuns 
\& Warren(1997)]{1997MNRAS.284L..11T} Theuns, T., \& Warren, S.~J.\ 1997, \mnras, 284, L11 


\bibitem[Thuan 
\& Gunn(1976)]{1976PASP...88..543T} Thuan, T.~X., \& Gunn, J.~E.\ 1976, \pasp, 88, 543 

\bibitem[Thuan 
\& Kormendy(1977)]{1977PASP...89..466T} Thuan, T.~X., \& Kormendy, J.\ 1977, \pasp, 89, 466 

%
%
%

\bibitem[Trentham 
\& Mobasher(1998)]{1998MNRAS.293...53T} Trentham, N., \& Mobasher, B.\ 1998, \mnras, 293, 53 

\bibitem[Uson et al.(1991)]{1991ApJ...369...46U} Uson, J.~M., Boughn, 
S.~P., \& Kuhn, J.~R.\ 1991, \apj, 369, 46

%
%
%
%
%

\bibitem[Vilchez-Gomez et 
al.(1994)]{1994A&A...283...37V} Vilchez-Gomez, R., Pello, R., \& Sanahuja, B.\ 1994, \aap, 283, 37 

\bibitem[Welch 
\& Sastry(1972)]{1972ApJ...171L..81W} Welch, G.~A., \& Sastry, G.~N.\ 1972, \apjl, 171, L81 

\bibitem[West(1993)]{1993MNRAS.265..755W} West, M.~J.\ 1993, \mnras, 265, 
755

\bibitem[West et al.(1995)]{1995ApJ...453L..77W} West, M.~J., Cote, P., 
Jones, C., Forman, W., \& Marzke, R.~O.\ 1995, \apjl, 453, L77

%
\bibitem[West et al.(2010)]{2010ApJ..} West, M.~J., et al.\ 2011, ApJ, accepted


%

\bibitem[Whiley et al.(2008)]{2008MNRAS.387.1253W} Whiley, I.~M., et al.\ 
2008, \mnras, 387, 1253

%
\bibitem[Williams et al.(2007)]{2007ApJ...654..835W} Williams, B.~F., et 
al.\ 2007, \apj, 654, 835

\bibitem[Woodley et al.(2010)]{2010ApJ...708.1335W} Woodley, K.~A., Harris, 
W.~E., Puzia, T.~H., G{\'o}mez, M., Harris, G.~L.~H., 
\& Geisler, D.\ 2010, \apj, 708, 1335

\bibitem[Yahagi 
\& Bekki(2005)]{2005MNRAS.364L..86Y} Yahagi, H., \& Bekki, K.\ 2005, \mnras, 364, L86 

\bibitem[Yang et al.(2009)]{2009ApJ...693..830Y} Yang, X., Mo, H.~J., 
\& van den Bosch, F.~C.\ 2009, \apj, 693, 830 

%

\bibitem[Zibetti et al.(2005)]{2005MNRAS.358..949Z} Zibetti, S., White, 
S.~D.~M., Schneider, D.~P., \& Brinkmann, J.\ 2005, \mnras, 358, 949 

\bibitem[Zwicky(1951)]{1951PASP...63...61Z} Zwicky, F.\ 1951, \pasp, 63, 61 

\end{thebibliography}
\end{document}